\documentclass[english,nofootinbib,notitlepage,showpacs,prd]{revtex4-1}
\pdfoutput=1
\usepackage[T1]{fontenc}
\usepackage[latin9]{inputenc}
\usepackage{color}
\usepackage{babel}
\usepackage{graphicx}
\usepackage{esint}
\usepackage[unicode=true,pdfusetitle,
 bookmarks=true,bookmarksnumbered=false,bookmarksopen=false,
 breaklinks=false,pdfborder={0 0 1},backref=false,colorlinks=false]
 {hyperref}

\makeatletter

\providecommand{\tabularnewline}{\\}

\@ifundefined{textcolor}{}
{%
 \definecolor{BLACK}{gray}{0}
 \definecolor{WHITE}{gray}{1}
 \definecolor{RED}{rgb}{1,0,0}
 \definecolor{GREEN}{rgb}{0,1,0}
 \definecolor{BLUE}{rgb}{0,0,1}
 \definecolor{CYAN}{cmyk}{1,0,0,0}
 \definecolor{MAGENTA}{cmyk}{0,1,0,0}
 \definecolor{YELLOW}{cmyk}{0,0,1,0}
}

\makeatother

\begin{document}

\title{Production of pentaquarks in $pA$-collisions}

\author{Iván Schmidt, M. Siddikov}

\affiliation{Departamento de Física, Universidad Técnica Federico Santa María,\\
 y Centro Científico - Tecnológico de Valparaíso, Casilla 110-V, Valparaíso,
Chile}

\preprint{USM-TH-XXX}


\begin{abstract}
We argue that a hidden-charm pentaquark recently observed in weak
decays of $\Lambda_{b}$ can be produced in proton-nucleus collisions
without electroweak intermediaries. We analyze the production cross-section
for several scenarios of internal structure and find that a cross-section
is sizable. This process can be studied both in collider as well
as in fixed-target experiments. In the former case, the pentaquarks
are produced at very forward rapidities, whereas in the latter case,
pentaquarks are produced with relatively small rapidities and can
be easily detected via invariant mass distribution of a forward $J/\psi$
and a comoving proton. Additionally, the suggested process allows to
check the existence of a neutral pentaquark $P_{c}^{0}$ (an isospin partner
of $P_{c}^{+}$) predicted in several models. The rapidity and transverse
momentum distributions of pentaquarks could provide comprehensive
information about the $\bar{c}c$ component of this exotic baryon. 
\end{abstract}
\maketitle

\section{Introduction}

\label{sec:Introduction}The recent discovery of hidden charm pentaquarks,
$P_{c}^{+}(4380)$ and $P_{c}^{+}(4450)$, in weak decays of $\Lambda_{b}$
hyperons~\cite{Aaij:2015tga} is without any doubts an important
step in the study of exotic baryons predicted by Gell-Mann~\cite{GellMann:1964nj}.
The existence of the $P_{c}^{+}$ pentaquark agrees with early observations
that there is an attractive Van der Waals like interaction between
$\bar{c}c$ and light quark matter, which could lead to  formation of bound
states~\cite{Brodsky:1997yr,Brodsky:1989jd} and could provide a natural
explanation for the large intrinsic charm of the proton suggested in \cite{Brodsky:1980pb,Brodsky:1981se}
as a phenomenological description of certain charm production processes.
As of now, the following facts have been established experimentally: the existence of
$P_{c}^{+}$, its mass, decay width and that $J/\psi\, p$ is one
of the possible decay channels. Albeit the latter fact might indicate
a large Fock component with $\bar{c}c$ in a singlet state, this does
not exclude other color-spin-flavour arrangements, like for example
a weakly bound state of one of the $D$-mesons and charmed $\Lambda_{c}/\Sigma_{c}$-hyperon
suggested in~\cite{Karliner:2015ina,Chen:2015moa,Wang:2015qlf,He:2015cea,Chen:2015loa,Scoccola:2015nia,Yang:2015bmv,Chen:2016heh},
a bound state of $\chi_{c}$ and $p$ suggested in~\cite{Meissner:2015mza},
a weakly bound state of $J/\psi$ and $p$~\cite{Kahana:2015tkb},
$\psi(2S)$ and $p$~\cite{Eides:2015dtr}, or a strongly bound system
of colored $\bar{c}c$ with light quarks~\cite{Mironov:2015ica}.
The formation of pentaquark $P_{c}^{+}$ in low-energy phenomenological
models indicates its existence~\cite{Xiao:2015fia,Meissner:2015mza},
albeit does not exclude a possibility that it is a kinematic effect
due to a Landau pole singularity in a triangle diagram~\cite{Guo:2015umn,Anisovich:2015xja}.
For this reason, it was concluded that a pentaquark existence could
be confirmed only after a study of additional decay channels~\cite{Wang:2015pcn},
as well as confirmation of the existence of other pentaquarks from $SU(3)$
flavor symmetry decuplet~\cite{Li:2015gta,Briceno:2015rlt,Feijoo:2015kts}. 

Another way to avoid the above-mentioned anomalous triangle singularity
is to study other production mechanisms and instead of weak $\Lambda_{b}$-decays
consider photoproduction of pentaquarks in $\gamma p$ collisions~\cite{Karliner:2015voa,Kubarovsky:2015aaa,Wang:2015jsa}
or $\pi p$ collisions~\cite{Lu:2015fva}. 

From a theoretical point of view, a crucial advantage of the $P_{c}^{+}$
compared to a putative $\Theta^{+}$ pentaquark (see~\cite{Hicks:2005gp}
for a review of $\Theta^{+}$ studies) is that it possesses two massive
$\bar{c}c$ quarks, whose dynamics can be described by perturbative
QCD methods. Additionally, in processes where $\bar{c}c$ is produced
diffractively, the typical distance between the quarks is small, $\sim1/m_{c}$,
and in certain cases there is an additional suppression $\sim\Lambda_{{\rm QCD}}/m_{c}$
due to a destructive interference of interaction amplitudes from $c$
and $\bar{c}$. This fact has been used extensively in the study of
charmonium and bottomonium production, where different models successfully
predict the cross-section~(see \cite{Brambilla:2004wf,Bedjidian:2004gd,Brodsky:2009cf}
for a review of the current experimental and theoretical situation). 

In this paper we suggest that $P_{c}^{+}$ might be produced in proton-nucleus
collisions in forward kinematics, as a two-stage process discussed
in the next Section~\ref{sec:Mechanism}. We assume that a $\bar{c}c$
pair needed for formation of a pentaquark is produced diffractively,
so we can apply the above-mentioned formalisms developed for charmonium
dynamics and study dynamics of a $\bar{c}c$ pair in $P_{c}^{+}$.
According to our estimates, this process has a sizable cross-section,
both when $\bar{c}c$ is in color singlet and in color octet states.

Potentially pentaquarks can be produced diffractively also in $pp$
collisions, when a $\bar{c}c$ pair from extrinsic charm after interaction
with the projectile leads to the formation of a $\bar{c}c$. However, typical
cross-sections of this process are smaller than in case of $pA$ collisions,
and production occurs at smaller rapidities. A full analysis of diffractive
pentaquark production in $pp$ collisions will be presented elsewhere. 

The paper is structured as follows. In Section~\ref{sec:Mechanism}
we discuss a suggested mechanism of pentaquark production and a framework
which we use for its description. For the sake of simplicity in this section we consider
proton-deuteron collisions, and take into account
only the contribution of extrinsic charm. In Section~\ref{sec:Nucl}
we generalize our framework for realistic large-$A$ nuclei. The contribution
of the intrinsic charm is discussed in Section~\ref{sec:intrinsic}.
In Section~\ref{sec:ModelWF} we discuss parametrizations of the
pentaquark and proton light-cone wave functions, and in Section~\ref{sec:NumericalResults}
we present numerical estimates for the pentaquark production cross-sections.

\section{Pentaquark production mechanisms}

\label{sec:Mechanism}

\begin{figure}
\includegraphics[scale=0.31]{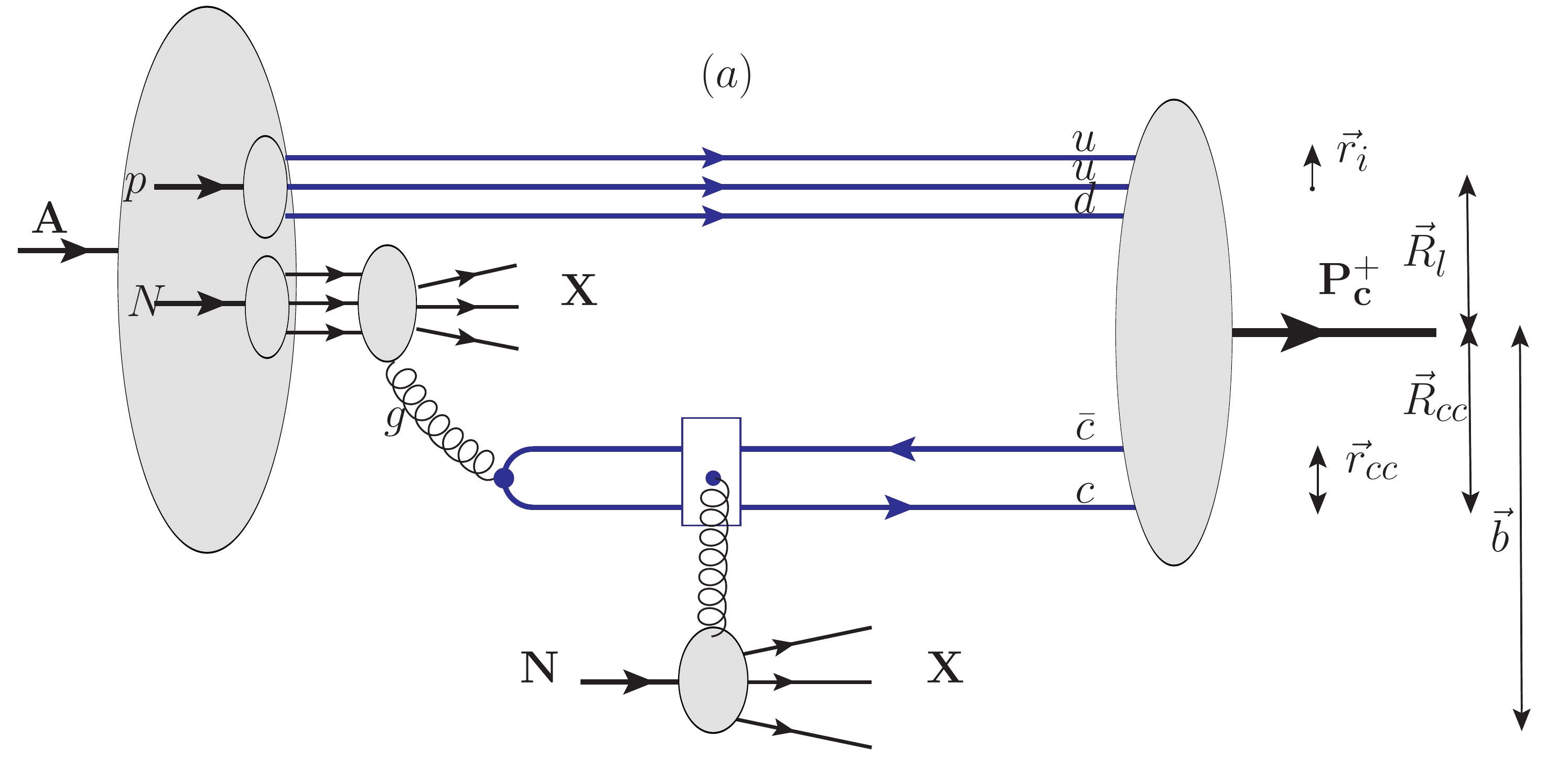}\includegraphics[scale=0.31]{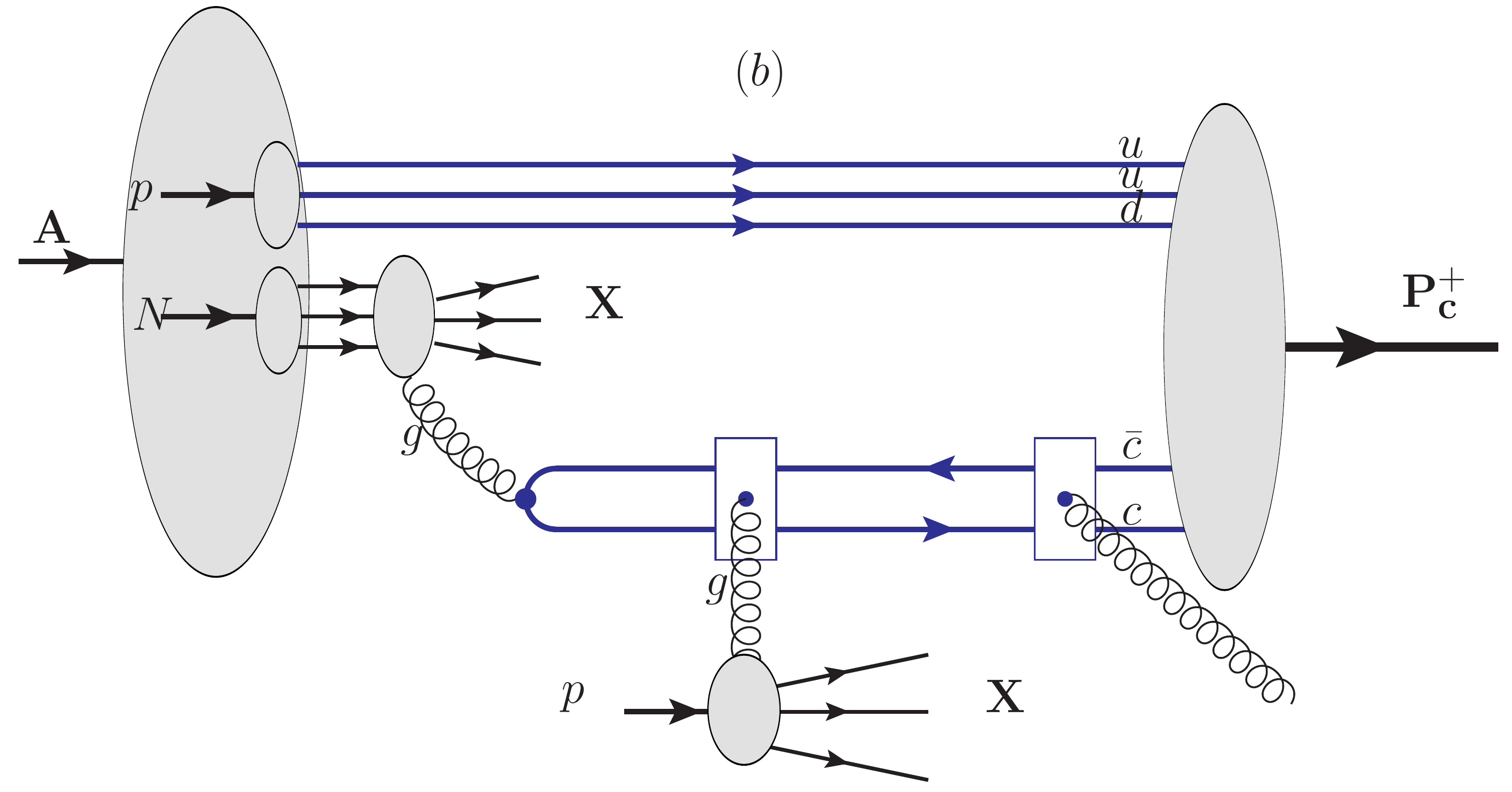}

\protect\caption{Perturbative diagrams contributing to the pentaquark production if
the $\bar{c}c$ pair is in color singlet state. Diagrams shown in
the left (right) pane probe $P$-wave ($S$-wave) component of the
$\bar{c}c$ dipole in a pentaquark. Each squared block implies a sum
of diagrams with a gluon attached to each of the quarks. In a diagram
(b) both emissions before and after interaction with a target are
possible. \label{fig:Diagrams_singlet} }
\end{figure}

In this section we analyze $P_{c}^{+}$ production in proton-deuteron
collisions at high energies, postponing a discussion of nuclear effects
until Section~\ref{sec:Nucl}. The dominant production mechanism
depends on the internal structure of this baryon. As was mentioned in
Section~\ref{sec:Introduction}, at this moment several competing
phenomenological models assume that a pentaquark is a bound state
of a hidden charm meson with a proton or a charmed $D$-meson and
$\Sigma_{c}/\Lambda_{c}$-hyperon. From the QCD point of view, the difference
between the two classes of models is in the color state of the $\bar{c}c$
state: the former case implies that the $\bar{c}c$ pair is in a color
singlet state, whereas in the latter case the $\bar{c}c$ pair is
mostly in the color octet state. However, we would like to note that
in QCD, if we assume dominance of a state with definite color or
spin of the $\bar{c}c$-component, due to exchange of virtual gluons between
light and heavy quarks this Fock state will certainly contain a small admixture
of all possible color-spin combinations. 

In the leading order over $\alpha_{s}\left(m_{c}\right)$, the pentaquark
production might proceed in a process schematically shown in a diagram
(a) of Fig~\ref{fig:Diagrams_singlet}. This process probes a color
singlet $\bar{c}c$ component of the pentaquark in a $P$-wave, as
suggested in~\cite{Meissner:2015mza}. A $\bar{c}c$-pair necessary
for a pentaquark formation is produced via a gluon splitting in $pA$
collisions. The produced $\bar{c}c$ has a negative invariant mass,
so in order to be able to produce a near-onshell $\bar{c}c$, it should
interact at least once with the target. To probe the $\bar{c}c$ pair
inside a pentaquark in a color singlet $S$-wave, the $\bar{c}c$
pair should emit at least one gluon (jet), as shown in the diagram
(b) of the Figure~\ref{fig:Diagrams_singlet}. 

\begin{figure}
\includegraphics[scale=0.31]{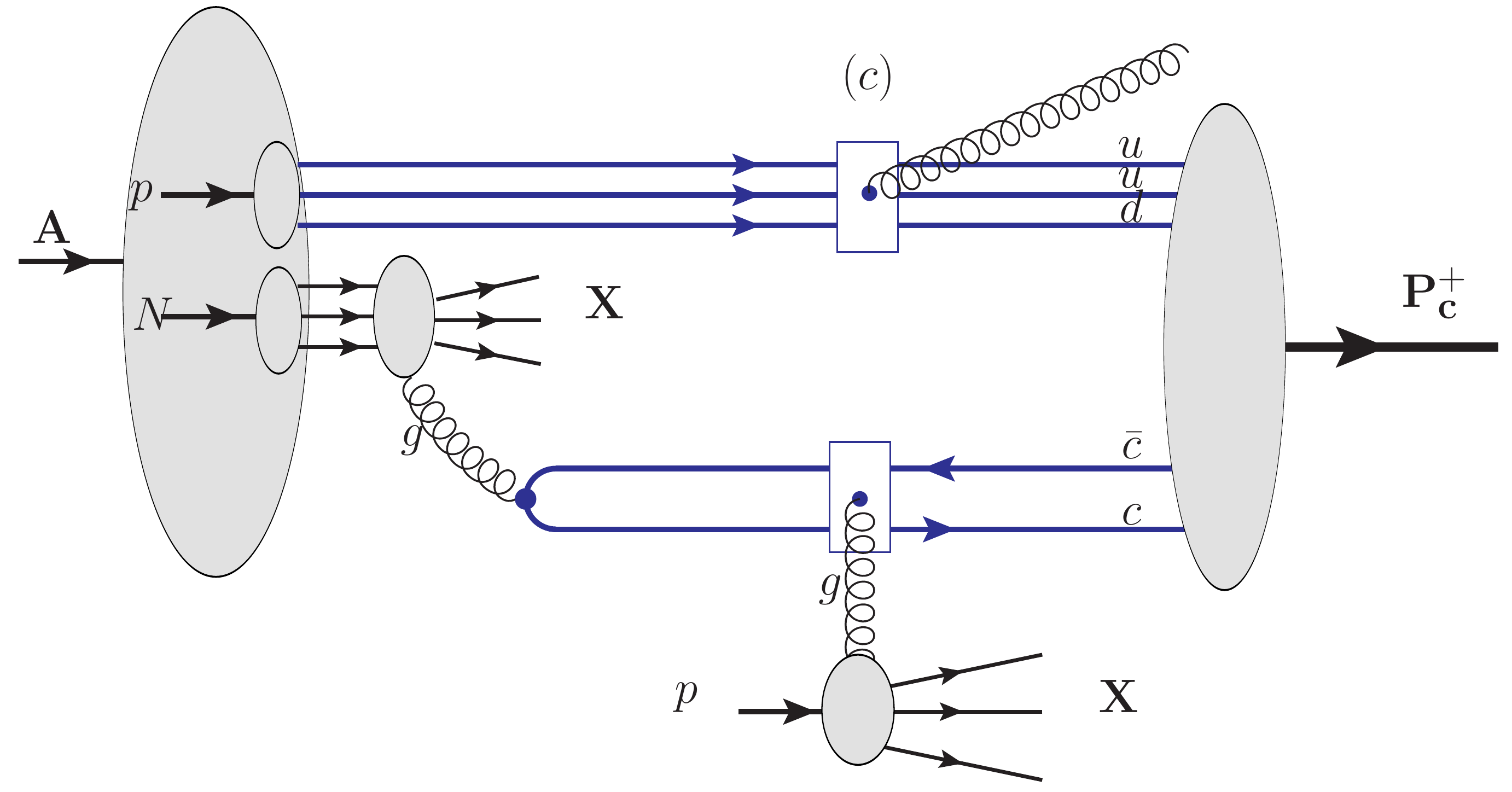}\includegraphics[scale=0.31]{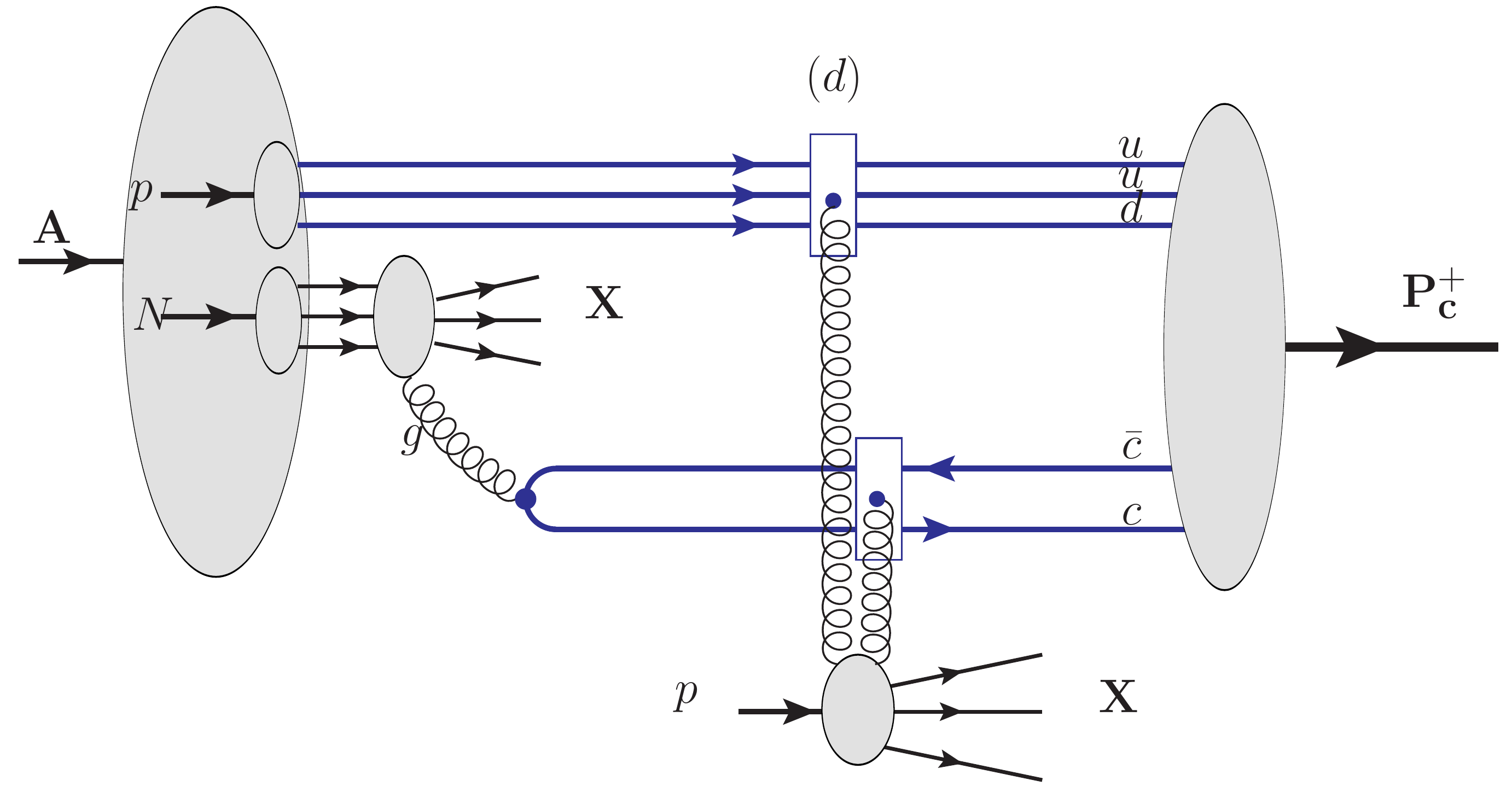}

\protect\caption{Perturbative diagrams contributing to the pentaquark production if
the $\bar{c}c$ pair is in color octet state. Both diagrams can probe
a $\bar{c}c$ dipole both in $S$- and $P$-wave. Each squared block
implies a sum of diagrams with a gluon attached to each of the quarks.
In case of $S$-wave production, there is an additional diagram with
$t$-channel gluon attached to a projectile gluon instead of $\bar{c}c$
(not shown). \label{fig:Diagrams_octet} }
\end{figure}

If the $\bar{c}c$ pair inside a pentaquark is in the color octet
state, as suggested in~\cite{Chen:2015moa,Wang:2015qlf,He:2015cea,Chen:2015loa,Scoccola:2015nia,Mironov:2015ica},
its production might proceed via processes shown in the Figure~\ref{fig:Diagrams_octet}.
In both cases the $\bar{c}c$ pair can be produced both in $S$- and
$P$-wave. 

In all four diagrams the typical values of a light-cone fraction $x_{1}$
carried by a projectile gluon from a proton cannot be very small:
otherwise light quarks and a $\bar{c}c$ pair separated by a large
rapidity gap cannot form a bound state. At the same time, typical
values of a variable $\langle x_{2}\rangle\sim M_{P_{c}}/x_{1}\sqrt{s}\ll1$.
In this kinematics, application of perturbative QCD might be
questionable due to saturation effects. For this reason, we adopt
a dipole model which naturally incorporates all saturation effects
and has been applied to the description of $\bar{c}c$ production in~\cite{Kopeliovich:2001ee,Kopeliovich:2005us,Kopeliovich:2015qna,Kopeliovich:2015vqa,Kopeliovich:2002yv}.
Since the produced dipoles are small and have a typical size $\langle r_{cc}\rangle\sim m_{c}^{-1}$,
up to $\mathcal{O}\left(1/m_{c}\right)$-corrections this approach
is equivalent to a $k_{T}$-factorization approach suggested in~\cite{Baranov:2007dw,Baranov:2015laa,Baranov:2015yea},
taking into account the relation of the dipole cross-section and the unintegrated
gluon PDF $\mathcal{F}\left(x,\, k_{\perp}\right)$ suggested in~\cite{GolecBiernat:1999ib}. 

The cross-section for the diagram $(a)$ in the Figure~\ref{fig:Diagrams_singlet}
has the form

\begin{eqnarray}
\frac{d\sigma^{(a)}}{dy} & = & \frac{1+x_{1}}{x_{1}}\, x_{1}g\left(x_{1}\right)\int\, d^{2}R_{cc}^{(1)}d^{2}R_{cc}^{(2)}\, d\alpha_{c}^{(1)}d^{2}r_{cc}^{(1)}\, d\alpha_{c}^{(2)}d^{2}r_{cc}^{(2)}\Phi_{\bar{c}c}^{\bar{\mu}\mu}\left(\alpha_{c}^{(1)},\,\vec{r}_{cc}^{(1)}\right)\Phi_{\bar{c}c}^{\bar{\nu}\nu*}\left(\alpha_{c}^{(2)},\,\vec{r}_{cc}^{(2)}\right)\label{eq:XSec_a}\\
 & \times & \Phi_{D}\left(-\frac{M_{P_{c}}}{M_{P_{c}}-2\, m_{c}}\vec{R}_{cc}^{(1)}\right)\Phi_{D}^{*}\left(-\frac{M_{P_{c}}}{M_{P_{c}}-2\, m_{c}}\vec{R}_{cc}^{(2)}\right)\mathcal{H}^{\bar{\mu}\mu}\left(\alpha_{c}^{(1)},\, x_{1},\,\vec{r}_{cc}^{(1)},\,\vec{R}_{cc}^{(1)}\right)\mathcal{H}^{\bar{\nu}\nu}\left(\alpha_{c}^{(2)},\, x_{1},\,\vec{r}_{cc}^{(2)},\,\vec{R}_{cc}^{(2)}\right)^{*}\nonumber \\
 & \times & \frac{1}{16}\left[\sigma\left(\alpha_{c}^{(1)}\vec{r}_{cc}^{(1)}+\bar{\alpha}_{c}^{(2)}\vec{r}_{cc}^{(2)}\right)+\sigma\left(\bar{\alpha}_{c}^{(1)}\vec{r}_{cc}^{(1)}+\alpha_{c}^{(2)}\vec{r}_{cc}^{(2)}\right)-\sigma\left(\alpha_{c}^{(1)}\vec{r}_{cc}^{(1)}-\alpha_{c}^{(2)}\vec{r}_{cc}^{(2)}\right)-\sigma\left(\bar{\alpha}_{c}^{(1)}\vec{r}_{cc}^{(1)}-\bar{\alpha}_{c}^{(2)}\vec{r}_{cc}^{(2)}\right)\right]\nonumber 
\end{eqnarray}
where $x_{1}$ is the light-cone fraction of the nucleon carried by
the gluon, $x_{1}g\left(x_{1}\right)$ is the gluon density in a projectile
proton, superscript indices 1 and 2 refer to the normal and complex
conjugate amplitudes, $\alpha_{c}$ and $\vec{r}_{cc}$ are the light-cone
fraction and  dipole size of the $c$-quarks in a $\bar{c}c$ pair, and
we also use variables $\vec{R}_{cc}$ for the distance between the center
of mass of the pentaquark and the $\bar{c}c$ pair, $\vec{r}_{i}$ for
the distance between the light quarks w.r.t. is center of mass ~%
\footnote{This set of variables is more convenient than the standard set of
Jacobi variables $\vec{r}_{cc}$, $\vec{R}_{cc}$,$\vec{\rho}_{1}$,$\vec{\rho}_{2}$,
where $\vec{\rho}_{1,2}$ are defined as
\begin{equation}
\vec{\rho}_{1}=\frac{2m_{c}\vec{R}_{cc}+m_{q}\vec{r}_{1}}{2m_{c}+m_{q}}-\vec{R}_{cc}=\frac{m_{q}}{2m_{c}+m_{q}}\left(\vec{r}_{1}-\vec{R}_{cc}\right),
\end{equation}
\begin{equation}
\vec{\rho}_{2}=\frac{(2m_{c}+m_{q})\vec{\rho}_{1}+m_{q}\vec{r}_{2}}{2m_{c}+2m_{q}}-\vec{\rho}_{1}=\frac{m_{q}}{2m_{c}+2m_{q}}\left(\vec{r}_{2}-\vec{\rho}_{1}\right),
\end{equation}
used for description of a classical many-body systems.%
}. The impact parameter $\vec{b}$ of the pentaquark does not appear
in the $p_{T}$-integrated cross-section.  The notation $\mathcal{H}^{\bar{\mu}\mu}$
in~\ref{eq:XSec_a} stands for the overlap of proton and pentaquark
wave functions, 
\begin{eqnarray}
\mathcal{H}^{\bar{\mu}\mu}\left(\alpha_{c},\,\xi,\,\vec{r}_{cc},\,\vec{R}_{cc}\right) & = & \int\prod_{i=1}^{3}\left(d\alpha_{i}dr_{i}\right)\delta^{2}\left(\sum_{i}\vec{r}_{i}\right)\delta\left(1-\sum_{i}\alpha_{i}\right)d\alpha_{c}\label{eq:HTilde}\\
 & \times & \Psi_{P_{c}}^{\dagger}\left(\frac{\alpha_{i}}{1+\xi},\vec{r}_{i}+\vec{R}_{l};\,\frac{\alpha_{c}\xi}{1+\xi},\vec{R}_{\bar{c}c}-\alpha_{c}\vec{r}_{\bar{c}c},\frac{\bar{\alpha}_{c}\xi}{1+\xi},\vec{R}_{\bar{c}c}+\bar{\alpha}_{c}\vec{r}_{\bar{c}c}\right)^{\nu_{1}\nu_{2}\nu_{3}\bar{\mu}\mu}\Psi_{p}^{\nu_{1}\nu_{2}\nu_{3}}\left(\alpha_{i},r_{i}\right),\nonumber 
\end{eqnarray}
where $\Psi_{p}$ and $\Psi_{P_{c}}$ are the light-cone wave functions
of the proton and pentaquark, $\bar{\mu}/\mu$ are spinor indices
of the $c$ and $\bar{c}$, and $\xi=\left(P_{c}^{+}-p^{+}\right)/p^{+}$
is the ratio of light-cone momenta of $\bar{c}c$ pair and light quarks.
For the wave function of heavy $\bar{c}c$ dipole we use the well-known
perturbative expression~\cite{Kogut:1969xa}

\begin{eqnarray}
\Phi_{\bar{c}c}^{\bar{\mu}\mu}\left(\alpha_{c},\, r\right) & = & \frac{\sqrt{\alpha_{s}}}{2\pi\sqrt{2}}\xi^{\mu}\left(m_{c}\sigma\cdot e+i\left(1-2\alpha_{c}\right)\left(\sigma\cdot n\right)\,\left(e\cdot\nabla\right)+\left[n\times e\right]\cdot\nabla\right)\xi^{\bar{\mu}}K_{0}\left(\epsilon r\right),\label{eq:Phi_cc}
\end{eqnarray}
where
\begin{eqnarray}
\epsilon & = & \sqrt{m_{c}^{2}-\alpha_{c}\left(1-\alpha_{c}\right)m_{G}^{2}},
\end{eqnarray}
and $m_{G}$ is the effective gluon mass. In the dipole cross-section
$\sigma(r)$ for the sake of brevity here and in what follows we suppress
the dependence on the Bjorken variable $x_{2}$, tacitly assuming that
a value $x_{2}\approx M_{cc}^{2}/(x_{1}\, s)$ is used. In (\ref{eq:XSec_a})
we assume that the light cone momentum of the nucleus is equally distributed
among the nucleons: as we will show in the next Section~\ref{sec:Nucl},
the light-cone distribution of nucleons is indeed very narrow. 

The term $\left(1+x_{1}\right)/x_{1}$ in the prefactor of~(\ref{eq:XSec_a})
has a pole at small $x_{1}$, which implies that the corresponding amplitude
should vanish in the limit $x_{1}\to0$ in order to provide a finite
integrated cross-section. As we will see below, if a realistic parametrization
is used for a pentaquark wave function, indeed such behavior is observed:
a $\bar{c}c$ pair separated by a large rapidity gap from the proton
cannot form a pentaquark.

The variable $x_{1}$ is related to the light-cone components of the pentaquark momentum $P_{c}$
in the nucleon-nucleon center of mass system as

\begin{equation}
P_{c}=\left(\left(1+x_{1}\right)\sqrt{s},\,\frac{M_{P_{c}}^{2}+P_{\perp}^{2}}{\left(1+x_{1}\right)\sqrt{s}},\,\vec{P}_{\perp}\right),\label{eq:Pc_momentum}
\end{equation}
where $\sqrt{s}$ is a collision energy per nucleon, $x_{1}$ is the
fraction of momentum carried by the gluon coming from one of the projectile
protons, $M_{P_{c}}$ is a pentaquark mass, for which we use a value
$M_{P_{c}}\approx4.4\,{\rm GeV}$ both for $P_{c}^{+}(4380)$ and
$P_{c}^{+}(4450)$, and $\vec{P}_{\perp}$ is the transverse momentum
of the produced pentaquark. From~\ref{eq:Pc_momentum} we can extract
a relation between the rapidity $y$ of the produced pentaquark and $x_{1}$.
This relation, in the nucleon-nucleon center of mass system, is given
by 
\begin{eqnarray}
y & = & \frac{1}{2}\ln\left(\frac{P_{c}^{+}}{P_{c}^{-}}\right)=\ln\left(\frac{\left(1+x_{1}\right)\sqrt{s}}{\sqrt{M_{P_{c}}^{2}+P_{\perp}^{2}}}\right),\label{eq:rapidity}
\end{eqnarray}
\emph{i.e.} in a collider experiments pentaquarks cannot be produced with rapidities smaller
than a threshold value 
\begin{equation}
y_{min}\left(s,\, P_{\perp}\right)=\ln\left(\frac{\sqrt{s}}{\sqrt{M_{P_{c}}^{2}+P_{\perp}^{2}}}\right),
\end{equation}
and are concentrated in narrow bins 
\begin{equation}
y\in\left(y_{min}\left(s,\, P_{\perp}\right),\, y_{min}\left(s,\, P_{\perp}\right)+\ln2\right).\label{eq:rapidity_bin}
\end{equation}
In the nucleus rest frame, relevant for fixed target experiments, the
results might be obtained by a shift 
\begin{equation}
y_{RF}=y-\ln\left(\frac{\sqrt{s}}{m_{N}}\right)=\ln\left(\frac{\left(1+x_{1}\right)m_{N}}{\sqrt{M_{P_{c}}^{2}+P_{\perp}^{2}}}\right).\label{eq:rapidity_RF}
\end{equation}

The cross-section corresponding to diagram $(b)$, which is the
dominant mechanism if the $\bar{c}c$ pair inside a pentaquark is in
$S$-wave, has the form 

\begin{eqnarray}
\frac{d\sigma^{(b)}}{dy} & = & \frac{5}{8}\, x_{1}g\left(x_{1}\right)\frac{1+x_{1}}{x_{1}}\int\, d^{2}R_{cc}^{(1)}d^{2}R_{cc}^{(2)}\,\, d^{2}\rho\, d\alpha_{G}d\alpha_{1}d^{2}r_{cc}^{(1)}\, d\alpha_{2}d^{2}r_{cc}^{(2)}\label{eq:XSec_b}\\
 & \times & \Phi_{D}\left(-\frac{M_{P_{c}}}{M_{P_{c}}-2\, m_{c}}\vec{R}_{cc}^{(1)}\right)\Phi_{D}^{*}\left(-\frac{M_{P_{c}}}{M_{P_{c}}-2\, m_{c}}\vec{R}_{cc}^{(2)}\right)\times\nonumber \\
 &  & \sum_{n,n'=1}^{6}\eta_{n}\eta_{n'}{\rm Tr}\left[\Lambda_{M}\Phi_{\bar{c}c}^{(n)}\left(\epsilon_{n},\,\vec{r}_{n}\right)\Phi_{cG}^{(n)}\left(\delta_{n},\,\vec{\rho}_{n}\right)\right]{\rm Tr}\left[\Lambda_{M}\Phi_{\bar{c}c}^{(n')}\left(\epsilon_{n'},\,\vec{r}_{n'}\right)\Phi_{cG}^{(n')}\left(\delta_{n'},\,\vec{\rho}_{n'}\right)\right],\nonumber \\
 & \times & \sigma\left(\vec{b}_{n}-\vec{b}_{n'}\right)\mathcal{H}^{\bar{\mu}\mu}\left(\frac{\alpha_{1}}{1-\alpha_{G}},\, x_{1}\left(1-\alpha_{G}\right),\,\vec{r}_{cc}^{(1)},\,\vec{R}_{cc}^{(1)}\right)\mathcal{H}^{\bar{\nu}\nu}\left(\frac{\alpha_{2}}{1-\alpha_{G}},\, x_{1}\left(1-\alpha_{G}\right),\,\vec{r}_{cc}^{(2)},\,\vec{R}_{cc}^{(2)}\right)\nonumber 
\end{eqnarray}
where in addition to the notations that appear in~(\ref{eq:XSec_a}) we introduced
$\alpha_{G}$ and $\vec{\rho}$ for the light-cone fraction and transverse
coordinate of the emitted gluon, $\alpha_{1,2}$ for the light-cone
fractions of the incident gluon momentum carried by the $c$-quark
in the amplitude and its conjugate. The gluon emission wave function
 $\Phi_{cg}$  in~(\ref{eq:XSec_b})  has the form 

\begin{eqnarray}
\Phi_{cg}\left(\beta,\rho\right) & \approx & \frac{i\sqrt{\alpha_{s}}}{\pi\sqrt{3}}\,\xi_{\mu}^{\dagger}\hat{\mathcal{D}}\xi_{\bar{\mu}}K_{0}\left(\delta\left|\vec{\rho}\right|\right),\label{eq:Phi_cg}
\end{eqnarray}
where
\begin{eqnarray}
\hat{\mathcal{D}} & = & 2\left(1-\frac{\beta_{1}}{2}\right)e_{f}\cdot\nabla+im_{c}\beta_{1}^{2}\left(n\times e_{f}\right)\cdot\sigma-i\beta_{1}\left(\nabla\times e_{f}\right)\cdot\sigma,\\
\delta & = & \sqrt{\beta^{2}m_{c}^{2}+\left(1-\beta\right)m_{G}^{2}},
\end{eqnarray}
 $e_{f}$ is a polarization vector of the emitted gluon, $\mu$ and
$\bar{\mu}$ are helicities of the quark before/after emission, and
$\beta$ is the ratio of the gluon light-cone fraction to the light
cone fraction of the initial quark. The vector $\vec{r}_{n},\,\vec{\rho}_{n},\,\vec{b}_{n}$,
as well as the coefficients $\epsilon,\,\delta$ in~(\ref{eq:Phi_cc},\ref{eq:Phi_cg})
are given by~%
\footnote{If we use a coordinate space evaluation, naively
we could expect that the coefficients $\epsilon$ for diagrams (5,6) should
fulfill $\epsilon_{5}=\epsilon_{1}$,~$\epsilon_{6}=\epsilon_{2}$,
but this is not so. This happens because in the evaluation of (\ref{eq:Phi_cc})
at least one of the quarks should be onshell, which is no longer true.
In the same way, offshellness of the final quark in the evaluation of diagrams
(3,4) leads to $\delta_{3}\not=\delta_{1}$ and $\delta_{4}\not=\delta_{2}$. %
} 
\begin{eqnarray}
\epsilon_{1}^{2} & = & \epsilon_{3}^{2}=m_{c}^{2}-\left(1-\alpha-\alpha_{G}\right)\left(\alpha+\alpha_{G}\right)\lambda^{2}\\
\delta_{1}^{2} & = & \delta_{5}^{2}=\alpha_{G}^{2}m_{c}^{2}+\alpha\left(\alpha+\alpha_{G}\right)\lambda^{2}\\
\delta_{3}^{2} & = & \epsilon_{5}^{2}=\frac{\left(\alpha+\alpha_{G}\right)\left(1-\alpha_{G}\right)\left(\alpha_{G}m_{c}^{2}+\alpha\left(1-\alpha-\alpha_{G}\right)\lambda^{2}\right)}{1-\alpha-\alpha_{G}}\\
\delta_{2}^{2} & = & \delta_{6}^{2}=\alpha_{G}^{2}m_{c}^{2}+\left(1-\alpha\right)\left(1-\alpha-\alpha_{G}\right)\lambda^{2}\\
\delta_{4}^{2} & = & \epsilon_{6}^{2}=\frac{\left(1-\alpha_{G}\right)\left(1-\alpha\right)\left(\alpha_{G}m_{c}^{2}+\alpha\left(1-\alpha-\alpha_{G}\right)\lambda^{2}\right)}{\alpha}\\
\epsilon_{2}^{2} & = & \epsilon_{4}^{2}=m_{c}^{2}-\left(1-\alpha\right)\alpha\lambda^{2}\\
\nonumber \\
\vec{r}_{1} & = & \vec{r}_{3}=\vec{r}_{5}=\frac{\alpha\vec{r}-\alpha_{G}\vec{\rho}}{\alpha+\alpha_{G}},\qquad\vec{\rho}_{1}=\vec{\rho}_{3}=\vec{\rho}_{5}=-\frac{\vec{\rho}+\left(1-\alpha-\alpha_{G}\right)\vec{r}}{\alpha+\alpha_{G}}\\
\vec{r}_{2} & = & \vec{r}_{4}=\vec{r}_{6}=-\frac{\left(1-\alpha-\alpha_{G}\right)\vec{r}+\alpha_{G}\vec{\rho}}{1-\alpha},\qquad\vec{\rho}_{2}=\vec{\rho}_{4}=\vec{\rho}_{6}=-\frac{\vec{\rho}-\alpha\vec{r}}{1-\alpha}\\
\vec{b}_{1} & = & \vec{b}+\frac{\alpha_{G}}{\alpha+\alpha_{G}}\vec{\rho}-\frac{\alpha(1-\alpha-\alpha_{G})}{\alpha+\alpha_{G}}\vec{r},\qquad\vec{b}_{2}=\vec{b}+\frac{\alpha_{G}}{1-\alpha}\vec{\rho}+\frac{\alpha(1-\alpha-\alpha_{G})}{1-\alpha}\vec{r},\\
\vec{b}_{3} & = & \vec{b}_{6}=\vec{b}-(1-\alpha-\alpha_{G})\vec{r},\qquad\vec{b}_{4}=\vec{b}_{5}=\vec{b}+\alpha\vec{r}.
\end{eqnarray}

and $\eta_{n}=\left\{ 1,1,-1,-1,-\alpha_{G},-\alpha_{G}\right\} $.

If the $\bar{c}c$ pair is in a color octet state, its production
may proceed via any of the mechanisms shown in the Figure~(\ref{fig:Diagrams_octet}).
The diagram ($c$) has the cross-section

\begin{eqnarray}
\frac{d\sigma^{(c)}}{dy} & = & \frac{1+x_{1}}{x_{1}}\, x_{1}g\left(x_{1}\right)\int d\alpha_{G}\int d^{2}\rho\int d^{2}R_{\bar{c}c}^{(1)}d^{2}R_{\bar{c}c}^{(2)}d\alpha_{c}^{(1)}d^{2}r_{cc}^{(1)}\, d\alpha_{c}^{(2)}d^{2}r_{cc}^{(2)}\Phi_{D}\left(-\frac{M_{P_{c}}}{M_{P_{c}}-2\, m_{c}}\vec{R}_{cc}^{(1)}\right)\label{eq:XSec_c}\\
 & \times & \Phi_{D}^{*}\left(-\frac{M_{P_{c}}}{M_{P_{c}}-2\, m_{c}}\vec{R}_{cc}^{(2)}\right)\Sigma^{(L)}\left(\alpha_{c}^{(1)},\, r_{cc}^{(1)},\alpha_{c}^{(2)},\, r_{cc}^{(2)}\right)\bar{\Psi}^{\bar{\mu}\mu*}\left(\alpha_{c}^{(1)},\, r_{cc}^{(1)}\right)\mathcal{O}_{P_{c}}^{\bar{\mu}\mu}\left(\alpha_{c}^{(1)},\,\alpha_{G},\, x_{1},\vec{\rho},\,\vec{R}_{\bar{c}c}^{(1)}\right)\nonumber \\
 & \times & \bar{\Psi}^{\bar{\nu}\nu}\left(\alpha_{c}^{(2)},\, r_{cc}^{(2)}\right)\mathcal{O}_{P_{c}}^{\bar{\nu}\nu*}\left(\alpha_{c}^{(2)},\,\alpha_{G},\, x_{1},\,\vec{\rho},\,\vec{R}_{\bar{c}c}^{(2)}\right),\nonumber 
\end{eqnarray}
where we introduced the shorthand notation $\Sigma^{(L)}$ for the cross-section
of the $\bar{c}c$ with internal orbital momentum $L$, 
\begin{eqnarray}
\Sigma^{(L=0)}\left(\alpha_{c}^{(1)},\, r_{cc}^{(1)},\alpha_{c}^{(2)},\, r_{cc}^{(2)}\right) & = & \frac{9}{16}\left[\sigma\left(\alpha_{c}^{(1)}\vec{r}_{cc}^{(1)}+\bar{\alpha}_{c}^{(2)}\vec{r}_{cc}^{(2)}\right)+\sigma\left(\bar{\alpha}_{c}^{(1)}\vec{r}_{cc}^{(1)}+\alpha_{c}^{(2)}\vec{r}_{cc}^{(2)}\right)-\sigma\left(\alpha_{c}^{(1)}\vec{r}_{cc}^{(1)}-\alpha_{c}^{(2)}\vec{r}_{cc}^{(2)}\right)\right.\\
 &  & \left.-\sigma\left(\bar{\alpha}_{c}^{(1)}\vec{r}_{cc}^{(1)}-\bar{\alpha}_{c}^{(2)}\vec{r}_{cc}^{(2)}\right)\right],\nonumber \\
\nonumber \\
\Sigma^{(L=1)}\left(\alpha_{c}^{(1)},\, r_{cc}^{(1)},\alpha_{c}^{(2)},\, r_{cc}^{(2)}\right) & = & \frac{5}{16}\left[\sigma\left(\alpha_{c}^{(1)}\vec{r}_{cc}^{(1)}+\bar{\alpha}_{c}^{(2)}\vec{r}_{cc}^{(2)}\right)+\sigma\left(\bar{\alpha}_{c}^{(1)}\vec{r}_{cc}^{(1)}+\alpha_{c}^{(2)}\vec{r}_{cc}^{(2)}\right)+\sigma\left(\alpha_{c}^{(1)}\vec{r}_{cc}^{(1)}-\alpha_{c}^{(2)}\vec{r}_{cc}^{(2)}\right)\right.\\
 &  & +\sigma\left(\bar{\alpha}_{c}^{(1)}\vec{r}_{cc}^{(1)}-\bar{\alpha}_{c}^{(2)}\vec{r}_{cc}^{(2)}\right)-2\sigma\left(\alpha_{c}^{(1)}\vec{r}_{cc}^{(1)}\right)-2\sigma\left(\bar{\alpha}_{c}^{(1)}\vec{r}_{cc}^{(1)}\right)-2\sigma\left(\alpha_{c}^{(2)}\vec{r}_{cc}^{(2)}\right)\nonumber \\
 &  & -\left.2\sigma\left(\bar{\alpha}_{c}^{(2)}\vec{r}_{cc}^{(2)}\right)\right],\nonumber 
\end{eqnarray}

and 
\begin{eqnarray}
\mathcal{O}_{P_{c}}^{\bar{\mu}\mu}\left(\alpha_{c},\,\alpha_{G},\,\xi,\,\vec{\rho},\,\vec{R}_{\bar{c}c}\right) & = & 3\int\prod_{k=1}^{3}\left(d\alpha_{k}d^{2}r_{k}\right)\delta^{2}\left(\sum_{k}\vec{r}_{k}\right)\delta\left(1-\sum_{k}\alpha_{k}\right)\Phi_{cG}\left(\frac{\vec{\rho}-\alpha_{G}\vec{r}_{1}}{1-\alpha_{1}}\right)\\
 & \times & \Psi_{P_{c}}^{\bar{\mu}\mu\dagger}\left(\left\{ \frac{\alpha_{k}-\alpha_{G}\delta_{k1}}{1+\xi-\alpha_{G}},\vec{r}_{k}+\vec{R}_{l}\right\} ;\,\frac{\alpha_{c}\xi}{1+\xi-\alpha_{G}},\,\vec{R}_{\bar{c}c}-\alpha_{c}\vec{r}_{\bar{c}c},\frac{\bar{\alpha}_{c}\xi}{1+\xi-\alpha_{G}},\,\vec{R}_{\bar{c}c}+\bar{\alpha}_{c}\vec{r}_{\bar{c}c}\right)\nonumber \\
 & \times & \psi_{p}\left(\left\{ \alpha_{k},\,\vec{r}_{k}\right\} \right)\nonumber 
\end{eqnarray}
for the overlap, numbering the active quark from which the emission takes
place with the index 1. As we will see below, this diagram is small since
the light quarks both in a proton and in a pentaquark are almost onshell.

\begin{figure}
\includegraphics[scale=0.3]{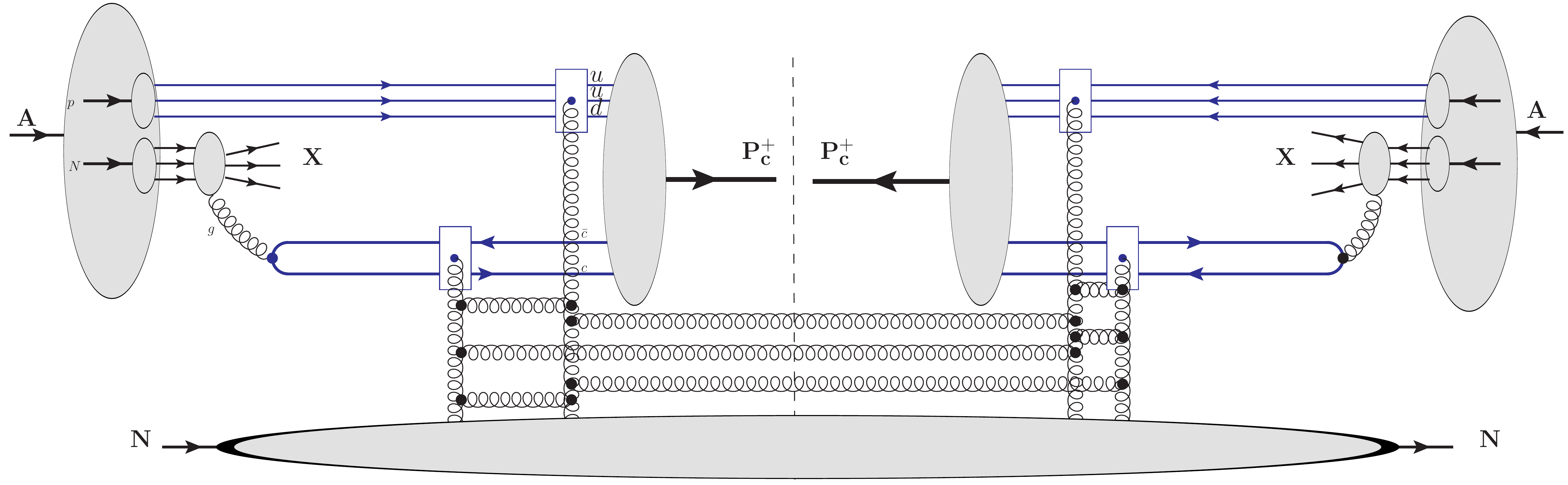}

\includegraphics[scale=0.3]{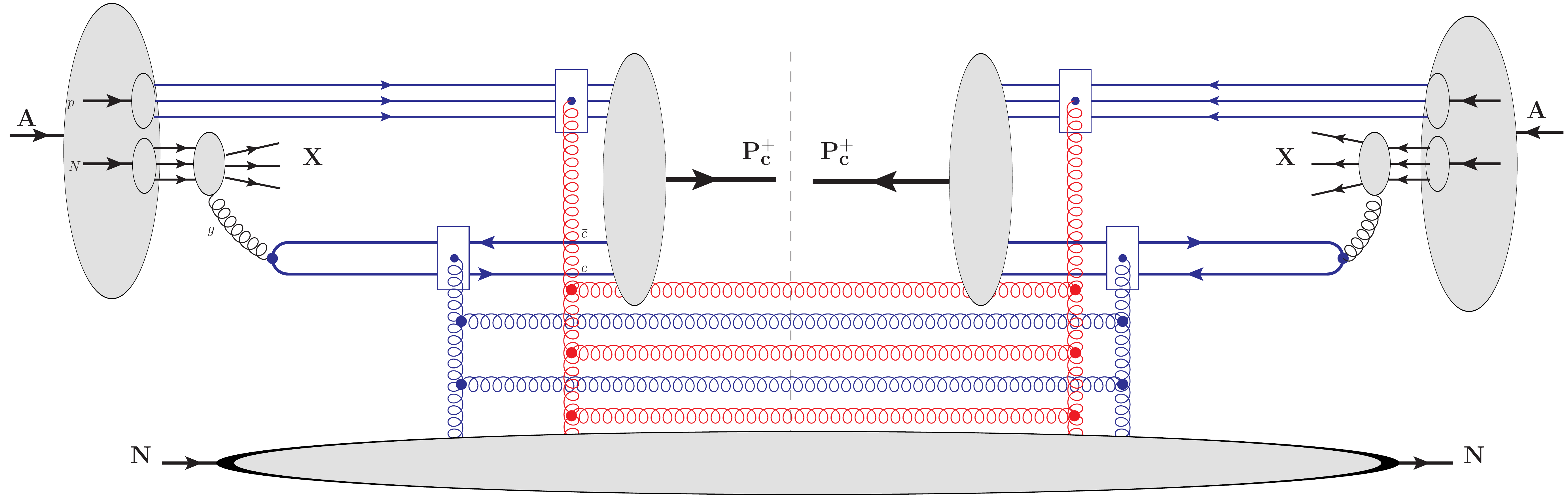}\protect\caption{\label{fig:Multipomeron}(color online) General multipomeron configurations
which contribute to a diagram (d) [upper plot] cannot be expressed
in terms of a dipole cross-section. However, the largest intercept
has a two-pomeron configuration [lower plot] (see the text for
explanations).}
\end{figure}

The evaluation of diagram $(d)$ in the general case is challenging, since
under reggeization potentially we may get multipomeron contributions
as shown in the upper part of the Figure~\ref{fig:Multipomeron}.
These contributions are not reducible to a mere dipole cross-section
and their evaluation presents a complicated problem. However, as was
shown in~\cite{Korchemsky:2001nx,ELBook}, in the large-$N_{c}$
limit the intercepts of these contributions are smaller, so the dominant
contribution at very high energies is a two-pomeron contribution shown
in the lower part of Figure~\ref{fig:Multipomeron}. In this limit
the corresponding cross-section is given by

\begin{eqnarray}
\frac{d\sigma^{(d)}}{dy} & = & 3\frac{1+x_{1}}{x_{1}}\, x_{1}g\left(x_{1}\right)\int d\alpha_{G}\int d^{2}\rho\int d^{2}R_{\bar{c}c}^{(1)}\int d^{2}R_{\bar{c}c}^{(2)}d\alpha_{c}^{(1)}d^{2}r_{cc}^{(1)}\, d\alpha_{c}^{(2)}d^{2}r_{cc}^{(2)}\label{eq:XSec_d}\\
 & \times & \Phi_{D}\left(-\frac{M_{P_{c}}}{M_{P_{c}}-2\, m_{c}}\vec{R}_{cc}^{(1)}\right)\Phi_{D}^{*}\left(-\frac{M_{P_{c}}}{M_{P_{c}}-2\, m_{c}}\vec{R}_{cc}^{(2)}\right)\nonumber \\
 & \times & \int d^{2}b\,\mathcal{N}\left(\alpha_{1}^{(1)}\vec{r}_{1}^{(1)}-\alpha_{1}^{(2)}\vec{r}_{1}^{(2)},\,\vec{b}+\vec{R}_{l}^{(12)}\right)\mathcal{S}{}^{(L)}\left(\alpha_{c}^{(1)},\, r_{cc}^{(1)},\alpha_{c}^{(2)},\, r_{cc}^{(2)},\,\vec{b}-\vec{R}_{cc}^{(12)}\right)\nonumber \\
 & \times & \bar{\Psi}^{\bar{\mu}\mu*}\left(\alpha_{c}^{(1)},\, r_{cc}^{(1)}\right)\bar{\Psi}^{\bar{\nu}\nu}\left(\alpha_{c}^{(2)},\, r_{cc}^{(2)}\right)\mathcal{O}_{P_{c}}^{\bar{\mu}\mu}\left(\alpha_{c}^{(1)},\,\alpha_{G},\, x_{1},\vec{\rho},\,\vec{R}_{\bar{c}c}^{(1)}\right)\mathcal{O}_{P_{c}}^{\bar{\nu}\nu*}\left(\alpha_{c}^{(2)},\,\alpha_{G},\, x_{1},\,\vec{\rho},\,\vec{R}_{\bar{c}c}^{(2)}\right),\nonumber 
\end{eqnarray}
where we introduced the notation

\begin{eqnarray}
\mathcal{S}{}^{(L=0)}\left(\alpha_{c}^{(1)},\, r_{cc}^{(1)},\alpha_{c}^{(2)},\, r_{cc}^{(2)},\,\vec{b}\right) & = & \frac{9}{16}\left[\mathcal{N}\left(\alpha_{c}^{(1)}\vec{r}_{cc}^{(1)}+\bar{\alpha}_{c}^{(2)}\vec{r}_{cc}^{(2)},\,\vec{b}\right)+\mathcal{N}\left(\bar{\alpha}_{c}^{(1)}\vec{r}_{cc}^{(1)}+\alpha_{c}^{(2)}\vec{r}_{cc}^{(2)},\,\vec{b}\right)\right.\\
 &  & \left.-\mathcal{N}\left(\alpha_{c}^{(1)}\vec{r}_{cc}^{(1)}-\alpha_{c}^{(2)}\vec{r}_{cc}^{(2)},\,\vec{b}\right)-\mathcal{N}\left(\bar{\alpha}_{c}^{(1)}\vec{r}_{cc}^{(1)}-\bar{\alpha}_{c}^{(2)}\vec{r}_{cc}^{(2)},\,\vec{b}\right)\right],\nonumber \\
\mathcal{S}^{(L=1)}\left(\alpha_{c}^{(1)},\, r_{cc}^{(1)},\alpha_{c}^{(2)},\, r_{cc}^{(2)},\,\vec{b}\right) & = & \frac{5}{16}\left[\mathcal{N}\left(\alpha_{c}^{(1)}\vec{r}_{cc}^{(1)}+\bar{\alpha}_{c}^{(2)}\vec{r}_{cc}^{(2)},\,\vec{b}\right)+\mathcal{N}\left(\bar{\alpha}_{c}^{(1)}\vec{r}_{cc}^{(1)}+\alpha_{c}^{(2)}\vec{r}_{cc}^{(2)},\,\vec{b}\right)\right.\\
 &  & +\mathcal{N}\left(\alpha_{c}^{(1)}\vec{r}_{cc}^{(1)}-\alpha_{c}^{(2)}\vec{r}_{cc}^{(2)},\,\vec{b}\right)+\mathcal{N}\left(\bar{\alpha}_{c}^{(1)}\vec{r}_{cc}^{(1)}-\bar{\alpha}_{c}^{(2)}\vec{r}_{cc}^{(2)},\,\vec{b}\right)\nonumber \\
 & - & \left.2\mathcal{N}\left(\alpha_{c}^{(1)}\vec{r}_{cc}^{(1)},\,\vec{b}\right)-2\mathcal{N}\left(\bar{\alpha}_{c}^{(1)}\vec{r}_{cc}^{(1)},\,\vec{b}\right)-2\mathcal{N}\left(\alpha_{c}^{(2)}\vec{r}_{cc}^{(2)},\,\vec{b}\right)-2\mathcal{N}\left(\bar{\alpha}_{c}^{(2)}\vec{r}_{cc}^{(2)},\,\vec{b}\right)\right],\nonumber 
\end{eqnarray}
where $L$ is the orbital momentum of internal motion of the $\bar{c}c$ pair,
and $\mathcal{N}\left(\vec{r},\,\vec{b}\right)$ is the dipole scattering
amplitude~\cite{Kowalski:2003hm} related to the dipole cross-section
as
\begin{equation}
\sigma(r)=\int d^{2}b\,\mathcal{N}\left(\vec{r},\,\vec{b}\right).
\end{equation}

\section{Effective Wave function from 2N correlations}

\label{sec:Nucl}In the previous Section~\ref{sec:Mechanism} we
considered pentaquark production in proton-deuteron collisions. In the
case of a heavy nucleus, we should replace the deuteron wave function
with a two-nucleon correlator. It was realized some time ago that this
object has properties similar to a deuteron wave function~\cite{Tagami:1959,Levinger:1979jf},
namely vanishes if the distance $R$ between the two nucleons is smaller
than 1 fm, and is suppressed at $R\gg2$~fm. A discussion of two-nucleon
correlations is usually done in terms of the two-particle density
\begin{equation}
\rho_{2}\left(\vec{x}_{1},\vec{x}_{2}\right)=A(A-1)\int\prod_{i=3}^{A}\left|\Psi_{A}\left(x_{1},...,x_{A}\right)\right|^{2}\approx A(A-1)\rho\left(\vec{x}_{1}\right)\rho\left(\vec{x}_{2}\right)\left(1-C\left(\vec{x}_{1},\vec{x}_{2}\right)\right),\label{eq:SRCdef}
\end{equation}
where $\rho\left(\vec{x}\right)$ is a normalized to unity (one-particle
nuclear density), and the function $C\left(\vec{x}_{1},\vec{x}_{2}\right)$
interpolates smoothly from 1 at distances smaller than 1 fm to 0 at
distances larger than 2-3 fm. This object has been extensively studied
in various theoretical models of nuclear structure~\cite{Alvioli:2009ab,Colle:2013nna,Viollier:1976ab,Cao:2012cv,Neff:2015xda},
yielding similar results. For  infinite nuclear matter, as well
as inside a finite nuclei far from a nuclear border, $C\left(\vec{x}_{1},\vec{x}_{2}\right)$
depends only on the distance between the two nucleons,
\begin{equation}
\left.C\left(\vec{x}_{1},\vec{x}_{2}\right)\right|_{\left|\vec{x}_{1,2}\right|\ll R_{A}}\approx C\left(r=\left|\vec{x}_{1}-\vec{x}_{2}\right|\right).
\end{equation}
 Experimentally, a nonzero value of the function $C$ reveals itself in
short-range correlations (SRC) of two nucleon pairs knocked out with
large momentum, $k>k_{F}$~\cite{Egiyan:2005hs,Frankfurt:1993sp},
where $k_{F}$ is a Fermi momentum. As was found in~\cite{Piasetzky:2006ai,Hen:2014nza,Colle:2015ena},
this effect depends on the isospin of the nucleons, and the dominant contribution
(>90\%) comes from $pn$ correlations. However, SRCs can appear not
only in specially crafted observables: as was demonstrated in~\cite{Alvioli:2008rw},
even the total quasielastic nucleon-nucleus cross-section is reduced
by 15\% due to SRCs.

From~(\ref{eq:SRCdef}) we may deduce a probability to find two nucleons
separated by a distance $r$, provided at least one of the nucleons
is a proton, 

\begin{equation}
\rho_{2N}\left(\vec{r}\right)\approx(A\, Z-1)\int d^{3}X\,\rho\left(\vec{X}-\frac{\vec{r}}{2}\right)\rho\left(\vec{X}+\frac{\vec{r}}{2}\right)\left(1-C\left(r\right)\right).
\end{equation}

\begin{figure}
\includegraphics{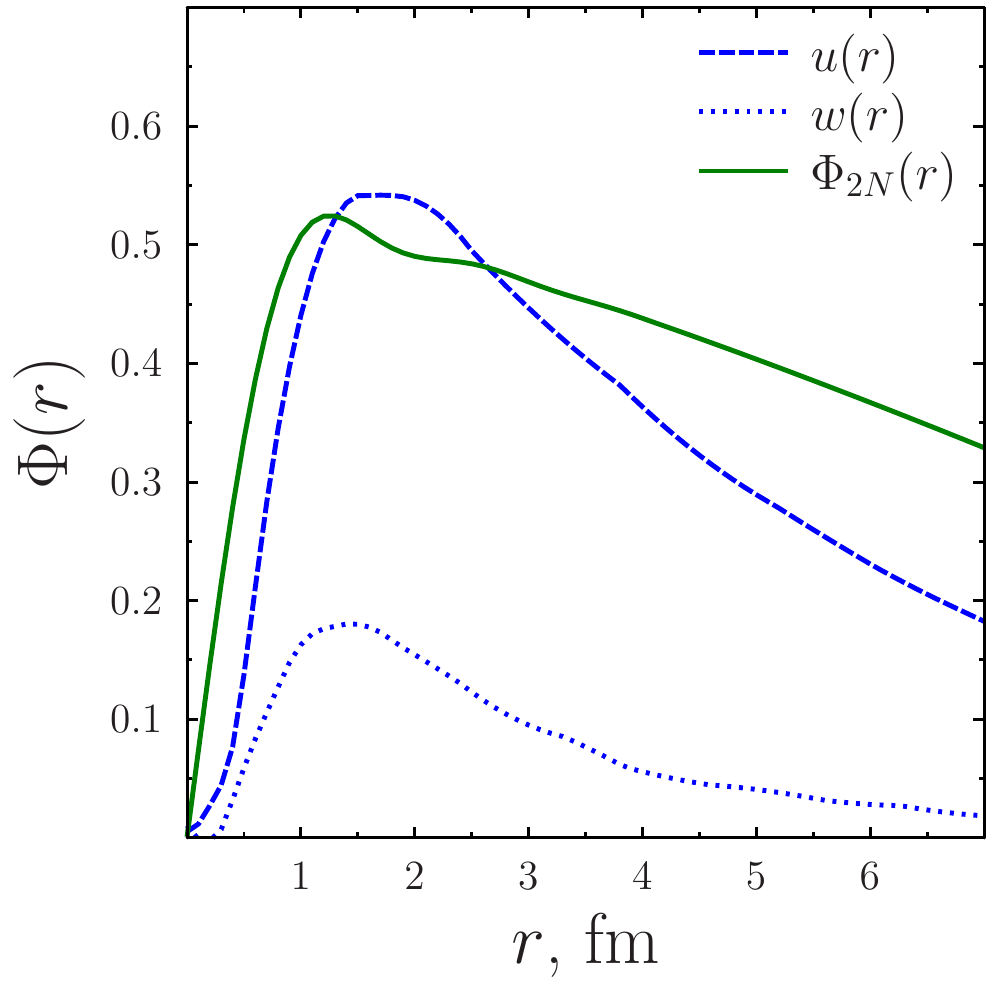}\protect\caption{\label{fig:WFs}(color online) Comparison of the effective 2-nucleon
wave function~$\Phi_{2N}$~(\ref{eq:WF2N}) (green solid lines)
with components of realistic deuteron wave function evaluated with
Argonne $v_{18}$ potential~\cite{ArgonneWF} (blue dashed lines).
The function $\Phi_{2N}$ is scaled by arbitrary factor to match the
peak value of the deuteron wave function component $u(r)$. }
\end{figure}

In what follows it is convenient to introduce an effective wave function
of relative motion of the $2N$ system, which we define as
\begin{equation}
\Phi_{2N}\left(\vec{r}\right)\equiv\rho_{2N}^{1/2}\left(\vec{r}\right)\approx\left[(A\, Z-1)\int d^{3}X\,\rho\left(\vec{X}-\frac{\vec{r}}{2}\right)\rho\left(\vec{X}+\frac{\vec{r}}{2}\right)\left(1-C\left(r\right)\right)\right]^{1/2},\label{eq:WF2N}
\end{equation}
and use a parametrization of $C(r)$ taken from~\cite{Alvioli:2009ab},
with a Woods-Saxon parametrization for nuclear densities. As one can
see from the Figure~\ref{fig:WFs}, the shape of the wave function
$\Phi_{2N}\left(\vec{r}\right)$ inside lead is very similar to
the realistic deuteron wave function evaluated with an Argonne $v_{18}$-potential~\cite{ArgonneWF}.
In our evaluations the wave function~(\ref{eq:WF2N}) contributes
in convolution with the pentaquark wave function, so the dominant contribution
stems from the region $r\sim2-3$~fm.

\begin{figure}
\includegraphics[scale=0.8]{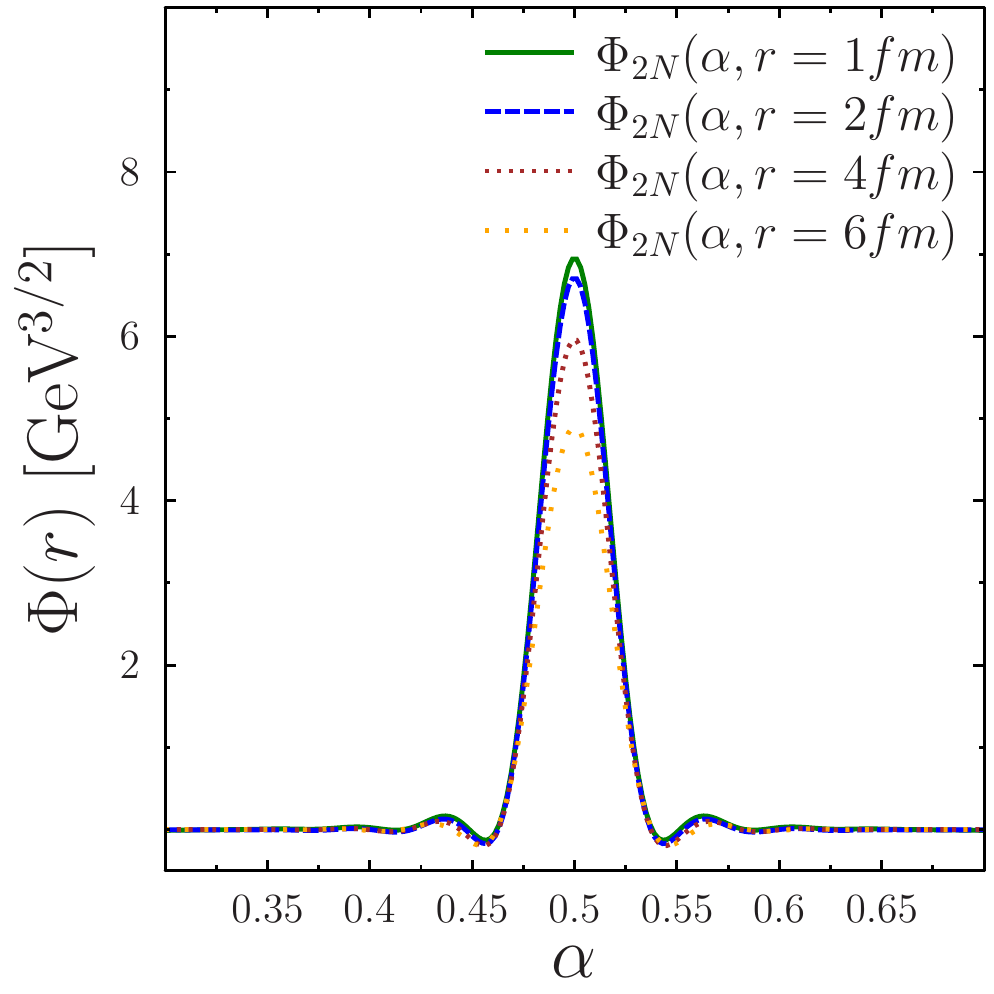}\includegraphics[scale=0.8]{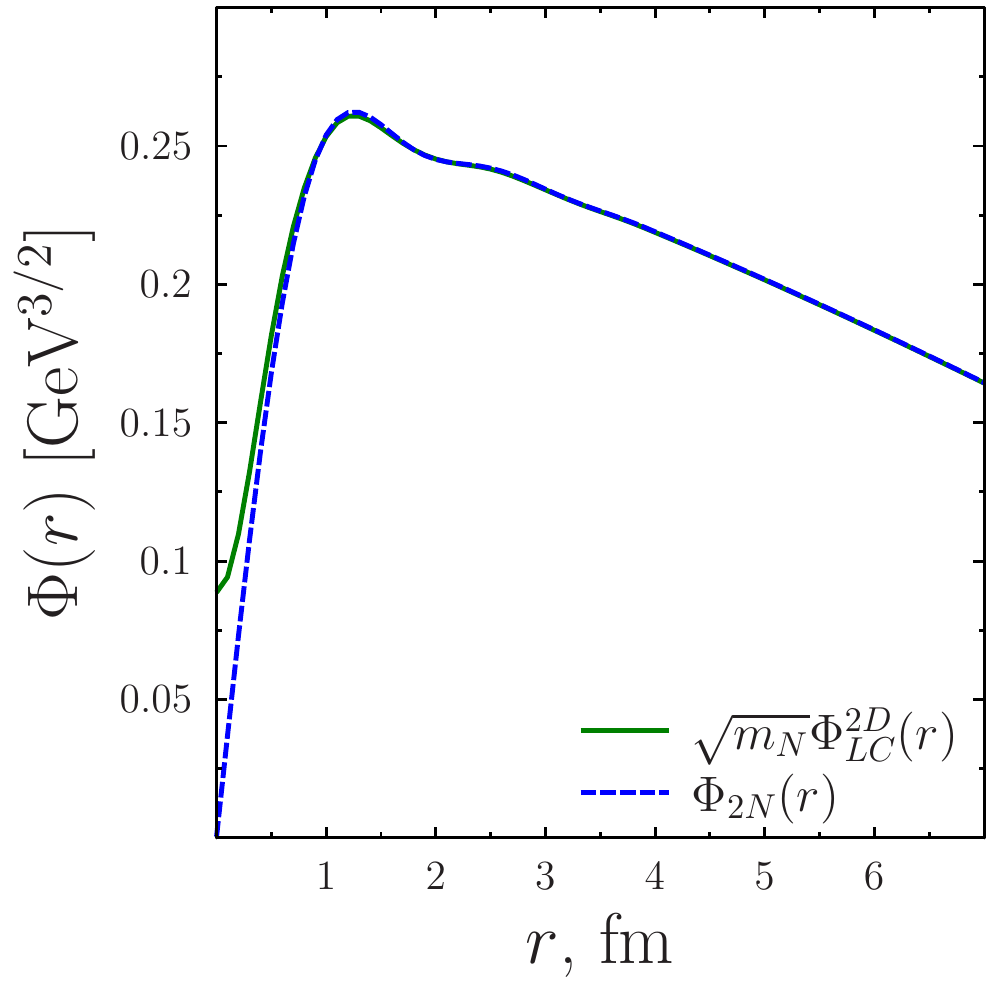}\protect\caption{\label{fig:WFs-1}(color online) Left: $\alpha$-dependence of the
light-cone correlator $\Phi_{2N}\left(\alpha,\, r_{\perp}\right)$
for several fixed values of $r$. As we can see, the peak is very
narrow. Right: Comparison of the light-cone wave function $\Phi_{LC}^{2D}\left(r_{\perp}\right)$
from~(\ref{eq:WF_2D}) (solid line) and 3D wave function~(\ref{eq:WF2N})
(dashed line). $\sqrt{m_{N}}$ is added to match dimensions.}
\end{figure}

The two-nucleon wave function~(\ref{eq:WF2N}) is given in the nucleus
rest frame, whereas in equations~(\ref{eq:XSec_a},\ref{eq:XSec_b},\ref{eq:XSec_c},\ref{eq:XSec_d})
we need a wave function in the light-cone formalism. In the general case a
relation between the two objects is a nontrivial dynamical problem,
however, under assumption that a motion of nucleons inside the nucleus
is nonrelativistic, we may use the relation suggested in~\cite{Frankfurt:1997fj,Hoyer:1999xe}
\begin{equation}
\tilde{\Phi}_{2N}\left(\alpha,k_{\perp}\right)=\sqrt[4]{\frac{k_{\perp}^{2}+m_{N}^{2}}{4\left[\alpha(1-\alpha)\right]^{3}}}\tilde{\Phi}_{2N}\left(k=\sqrt{\frac{k_{\perp}^{2}+(2\alpha-1)^{2}m_{N}^{2}}{4\alpha(1-\alpha)}}\right),\label{eq:Frankfurt}
\end{equation}
where $\tilde{\Phi}_{2N}$ stands for a Fourier transform of the corresponding
wave function. As follows from~(\ref{eq:Frankfurt}) and is seen
from the left pane of the Figure~\ref{fig:WFs-1} , the $\alpha$-dependence
of the wave function is strongly peaked near a value $\alpha\approx1/2$,
for this reason in Equations~(\ref{eq:XSec_a},\ref{eq:XSec_b},\ref{eq:XSec_c},\ref{eq:XSec_d})
we do not write a convolution over a light-cone fraction $\alpha$
carried by a nucleon, tacitly assuming that for all physical observables
\begin{equation}
\int d\alpha\,\Phi_{2N}\left(\alpha,\, r_{\perp}\right)\mathcal{A}\left(\alpha,...\right)\approx\int d\alpha\,\Phi_{2N}\left(\alpha,\, r_{\perp}\right)\mathcal{A}\left(\alpha\approx\frac{1}{2},\,...\right)+\mathcal{O}\left(\left\langle \left(\alpha-\frac{1}{2}\right)^{2}\right\rangle \right),\label{eq:WF_Phys}
\end{equation}
\emph{i.e.} in Equations~(\ref{eq:XSec_a},\ref{eq:XSec_b},\ref{eq:XSec_c},\ref{eq:XSec_d})
we should replace the deuteron wave function~$\Phi_{D}$ with 
\begin{equation}
\Phi_{LC}^{2D}\left(r_{\perp}\right)=\int d\alpha\,\Phi_{2N}\left(\alpha,\, r_{\perp}\right)\label{eq:WF_2D}
\end{equation}

In the right pane of the Figure~\ref{fig:WFs-1} we compare the ~$\Phi_{LC}^{2D}\left(r_{\perp}\right)$
with a rest frame wave function~(\ref{eq:WF2N}). As we can see,
the difference between the two objects is relevant only at small-$r$.

\section{On intrinsic charm contribution}

\label{sec:intrinsic}In addition to the extrinsic charm contribution
discussed in a previous Section~\ref{sec:Mechanism}, an additional
contribution might be due to intrinsic charm of a projectile proton,
first suggested in~\cite{Brodsky:1980pb,Brodsky:1981se}. This intrinsic
charm is known with considerable uncertainty (see e.g.~\cite{Jimenez-Delgado:2014zga}
for a short review). Its contribution to forward pentaquark production
might be obtained by a simple replacement of extrinsic charm with
intrinsic charm wave function in~(\ref{eq:XSec_a},\ref{eq:XSec_b},\ref{eq:XSec_c},\ref{eq:XSec_d}),
\begin{equation}
x_{1}g\left(x_{1}\right)\bar{\Psi}^{\bar{\mu}\mu*}\left(\alpha_{c},\, r_{cc}\right)\Phi_{D}\left(-\frac{M_{P_{c}}}{M_{P_{c}}-2\, m_{c}}\vec{R}_{cc}\right)\Psi_{N}^{\nu_{1}\nu_{2}\nu_{3}}\left(\alpha_{i},\vec{r}_{i}\right)\to\sqrt{P_{5}}\,\Psi_{{\rm IC}}^{\bar{\mu}\mu\nu_{1}\nu_{2}\nu_{3}}\left(\alpha_{i},\vec{r}_{i}\right),
\end{equation}
 where $P_{5}$ is a normalization coefficient which takes into account
the amount of intrinsic charm inside a proton, and $\Psi_{{\rm IC}}$
is the (unknown) wave function of the $uudc\bar{c}$ Fock component.
Usually the amount of intrinsic charm is quantized in terms of the
fraction of the momentum of a proton carried by $\bar{c}c$, and current
phenomenological estimates for this quantity indicate a very small
amount, $\langle x\rangle_{{\rm IC}}\sim0.15-0.5\%$~\cite{Jimenez-Delgado:2014zga}.
In view of this, in what follows we neglect the contribution of the
intrinsic charm, tacitly assuming that its contribution could only
increase the extrinsic charm cross-sections. Besides, in case of $pA$
collisions we expect that the extrinsic charm cross-sections will
be enhanced by a factor $\sim A^{1/3}$ compared to the intrinsic
charm contribution.

\section{Parametrization of pentaquark and nucleon wave function}

\label{sec:ModelWF}While there are very detailed theoretical models
for internal structure of a putative $\Theta^{+}$(see e.g.~\cite{Diakonov:1997mm,Lorce:2006nq}),
as of now there is no parameterizations of the light-cone wave functions
of the $P_{c}$-pentaquark known from the literature. From phenomenological
models it is expected that a pentaquark $P_{c}$ could be a ``molecular''
state of $J/\psi\, p$, $\chi_{c}\, p\,$ $\Sigma_{c}\bar{D}$ or
$\Lambda_{c}\bar{D}$. This implies that a $\bar{c}c$ can be predominantly
either in a color singlet or in a color octet state. In the general case,
wave functions obtained in effective models are in the rest frame
of the baryon. Boosting them to a light-cone frame presents a complicated
dynamical problem and mixes Fock components with different number
of quarks. For this reason, a direct modeling might be a better approach.
For this pioneering study, we completely disregard the spin structure
of light quarks both in the pentaquark and in the proton, tacitly assuming
that it is the same. We assume that the proton wave function in its rest
frame has the form~\cite{Kopeliovich:2005us}\\
\begin{equation}
\Psi_{p}\left(\left\{ \alpha_{i},\vec{r}_{i}\right\} \right)=f_{3}\left(\alpha_{1},\alpha_{2},\alpha_{3}\right)\left.\frac{\sqrt{3}\,}{\pi R_{p}^{2}}\exp\left(-\frac{1}{2\, R_{p}^{2}}\left(r_{1}^{2}+r_{2}^{2}+r_{3}^{2}\right)\right)\right|_{\sum_{i}\vec{r}_{i}=0},\label{eq:proton_Gauss}
\end{equation}
where the parameter $a$ is fixed from the charge radius of the proton
$r_{ch}^{2}$ and equals $a=3/\left(2r_{ch}^{2}\right)$. For modelling
the light-cone dependence of a pentaquark, we follow the general receipt
suggested long ago in~\cite{Brodsky:1980pb,Brodsky:1981se} for a
baryon with $n$ constituents and choose
\begin{equation}
f_{n}\left(\alpha_{1},...,\alpha_{n}\right)=\left.\frac{N_{n}}{\left(M_{B}^{2}-\sum_{i=1}^{n}\frac{m_{i}^{2}}{\alpha_{i}}\right)}\right|_{\sum\alpha_{i}=1},\label{eq:Brodsky_IC}
\end{equation}
where $M_{B}$ is a mass of a baryon and $m_{i}$ is the mass of constituent
quarks. This function has a smooth behavior provided $M_{B}<\sum m_{i}$.
For the masses of light and heavy quarks in what follows we take $m_{l}\approx0.35$~GeV
and $m_{c}\approx1.8$~GeV respectively. The normalization constant
$N_{n}$ is fixed from the normalization condition
\begin{equation}
\int\prod_{i}d\alpha_{i}\delta\left(1-\sum_{i}\alpha_{i}\right)\left|f\left(\left\{ \alpha_{i}\right\} \right)\right|^{2}=1.\label{eq:norm}
\end{equation}

As was discussed in~\cite{Brodsky:1980pb,Brodsky:1981se}, this function
provides a correct endpoint behavior of parton PDFs near $\alpha_{i}\approx1$.
Also, it vanishes when a light-cone fraction of any parton goes to
zero, which guarantees that the total cross-section is finite, as was
discussed in the previous Section~\ref{sec:Mechanism}. 

\begin{figure}
\includegraphics[scale=0.4]{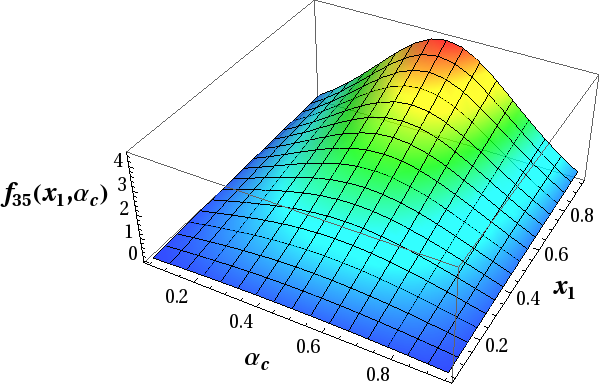}\protect\caption{\label{fig:f35}(color online) Dependence of the overlap of proton
and pentaquark wave functions on $x_{1}$ and $\alpha_{c}$.}
\end{figure}

For the pentaquark, modeling is a bit more complicated due to a myriads
of possible spin-orbital arrangements of all quarks. For us it is
important to distinguish the case when $\bar{c}c$ pair is either
in $S$-wave or in a $P$-wave. In what follows, we assume that it
has a structure similar to~(\ref{eq:proton_Gauss}),
\begin{eqnarray}
\Psi_{P_{c}}\left(\left\{ \alpha_{i},\vec{r}_{i}\right\} \right) & = & f_{5}\left(\alpha_{i}\right)v^{\bar{\mu}}\left(p_{\bar{c}}\right)u^{\mu}\left(p_{c}\right)\left.\frac{\sqrt{3}\,}{\pi R_{p}^{2}}\exp\left(-\frac{1}{2\, R_{p}^{2}}\left(r_{1}^{2}+r_{2}^{2}+r_{3}^{2}\right)\right)\right|_{\sum_{i}\vec{r}_{i}=0}\times\label{eq:Pc_WF-sym}\\
 & \times & \frac{\hat{U}\left(\vec{r}_{cc}\right)}{\pi\left\langle r_{cc}\right\rangle \left\langle R_{cc}\right\rangle }\exp\left(-\frac{r_{\bar{c}c}^{2}}{2\left\langle r_{cc}^{2}\right\rangle }-\frac{R_{\bar{c}c}^{2}}{2\left\langle R_{cc}^{2}\right\rangle }\right),\nonumber \\
\hat{U}\left(\vec{r}_{cc}\right) & = & \left\{ \begin{array}{cc}
1 & \bar{c}c=S-{\rm wave,\, octet}\\
\mathcal{N}_{2S}\left(r_{\bar{c}c}-a_{2S}\right) & S-{\rm wave,\, singlet}\\
\left(\frac{x_{\bar{c}c}\pm iy_{\bar{c}c}}{\sqrt{2}\left\langle r_{cc}\right\rangle }\right) & P-{\rm wave}
\end{array}\right.,
\end{eqnarray}
where $x_{\bar{c}c}$, $y_{\bar{c}c}$ are the components of the vector
$r_{\bar{c}c}$, $v^{\bar{\mu}}$ and $u^{\mu}$ are spinors which
correspond to the $\bar{c}$ and $c$-quark~%
\footnote{The parametrization~(\ref{eq:Pc_WF-sym}) assumes that the $\bar{c}c$
pair in a pentaquark can have any spin with equal probability. In the
case of models with internal structure additional spin projectors
should be added. We expect the uncertainty due to spin factors to result
in a factor of two uncertainty in the cross-section. %
}, and $f_{5}$ is given by~(\ref{eq:Brodsky_IC}). For the color
singlet $S$-wave we assume that the wave function has a node~%
\footnote{If we assume that the $\bar{c}c$ pair inside a $P_{c}^{+}$ is in
a color singlet $1S$ state and does not have any nodes, it would
be challenging to explain a relatively narrow width of its decay into
$J/\psi$ and $p$.%
}, i.e. it is a bound state of $\psi(2S)$ and a proton, as suggested
in~\cite{Eides:2015dtr}. The node position $a_{2S}=\left\langle r_{cc}\right\rangle \sqrt{\pi}$/
$2$ and the value of a normalization constant $\mathcal{N}_{2S}=2/\left(\left\langle r_{cc}\right\rangle \sqrt{4-\pi}\right)$
may be extracted from orthonormality with the 1S state and are in reasonable
agreement with numerical solutions from potential models of charmonium~\cite{Eichten:1979ms},

For a singlet $\bar{c}c$, the relevant cross-section is controlled
by an overlap~(\ref{eq:HTilde}), which in this case takes the form
\begin{eqnarray}
\mathcal{H}^{\bar{\mu}\mu}\left(\alpha_{c},\,\xi,\,\vec{r}_{cc},\,\vec{R}_{\bar{c}c}\right) & = & \frac{\hat{U}\left(\vec{r}_{cc}\right)}{\pi\left\langle r_{cc}\right\rangle \left\langle R_{cc}\right\rangle }v^{\bar{\mu}}\left(p_{\bar{c}}\right)u^{\mu}\left(p_{c}\right)f_{3,5}\left(\xi,\,\alpha_{c}\right)\exp\left(-\frac{r_{\bar{c}c}^{2}}{2\left\langle r_{cc}^{2}\right\rangle }-\frac{R_{\bar{c}c}^{2}}{2\left\langle R_{cc}^{2}\right\rangle }\right),\label{eq:Hmumu_overlap-1-1}
\end{eqnarray}
and we introduced a shorthand notation 
\begin{eqnarray}
f_{3,5}\left(\xi,\,\alpha_{c}\right) & \equiv & \int\prod_{i=1}^{3}d\alpha_{i}f_{5}\left(\left\{ \frac{\alpha_{i}}{1+\xi}\right\} _{{\rm light}},\,\frac{\xi\alpha_{c}}{1+\xi},\,\frac{\xi\bar{\alpha}_{c}}{1+\xi}\right)f_{3}\left(\alpha_{i}\right)\delta\left(\sum_{i}\alpha_{i}=1\right).
\end{eqnarray}
The dependence of the overlap $f_{3,5}\left(\xi,\,\alpha_{c}\right)$
is shown in Figure~\ref{fig:f35}. For very small-$x$, it vanishes
roughly as $f_{3,5}\left(\xi,\,\alpha_{c}\right)\sim\xi$. Physically,
this means that it is not possible to form a pentaquark if the $\bar{c}c$
is separated by a large rapidity gap from the proton. As a function
of $\alpha_{c}$, the distribution is close to the asymptotic form $\sim\alpha_{c}\left(1-\alpha_{c}\right)$. 

\section{Numerical results and discussion}

\label{sec:NumericalResults}

In this paper we consider only the $P_{\perp}$-integrated cross-section.
Since nothing is known about the light-cone wave function of the pentaquark,
in what follows we fix the values of parameters in the pentaquark wave
function~\ref{eq:Pc_WF-sym} as 
\begin{equation}
\left\langle R_{cc}^{2}\right\rangle \approx\frac{2}{3}\left\langle R_{p}^{2}\right\rangle \approx(0.98\,{\rm fm})^{2},\quad\left\langle r_{cc}^{2}\right\rangle \approx\frac{2}{3}\left\langle r_{J/\psi}^{2}\right\rangle \approx(0.33\,{\rm fm})^{2},\quad m_{c}\approx1.5\,{\rm GeV.}\label{eq:params}
\end{equation}

The value of $\left\langle R_{cc}\right\rangle $ in~(\ref{eq:params})
corresponds to the average distance between a proton and a $\bar{c}c$
pair, $\left\langle R_{P_{c}}\right\rangle \approx M_{P_{c}}/(M_{P_{c}}-2m_{c})\approx3\,{\rm fm}$
in agreement with a pentaquark-as-molecule picture~\cite{Chen:2015moa,Wang:2015qlf,Meissner:2015mza,He:2015cea,Chen:2015loa,Scoccola:2015nia,Yang:2015bmv,Eides:2015dtr,Kahana:2015tkb}.

For the dipole cross-section $\sigma(r)$, we use a GBW parametrization~\cite{GolecBiernat:1999ib}.
The key virtue of this parametrization is that due to its simplicity
it allows to simplify some integrals. For typical values of $\left\langle r_{cc}\right\rangle \sim2/m_{c}$
this parametrization differs from more accurate IP-Sat fit~\cite{Rezaeian:2012ji}
within 10-20\%, much less than an uncertainty due to pentaquark wave
function. For the $b$-dependent cross-section $\mathcal{N}\left(\vec{r},\,\vec{b}\right)$
in (\ref{eq:XSec_d}) we take the simple factorized form 
\begin{eqnarray}
\mathcal{N}\left(\vec{r},\,\vec{b}\right) & = & \sigma(r)T(b),\\
T(b) & = & \frac{1}{2\pi B_{D}}\exp\left(-\frac{b^{2}}{2B_{D}}\right),
\end{eqnarray}
with $B_{D}=4$~GeV$^{-2}$. This simplification allows to evaluate
analytically integrals over transverse coordinates in~(\ref{eq:XSec_d}).
For heavy dipoles of size $\left\langle r_{cc}\right\rangle \sim2/m_{c}$
this approximation differs from a more elaborated IP-Sat fit~\cite{Rezaeian:2012ji}
by small $\sim1/m_{c}^{2}$ corrections. 

Our estimates of pentaquark cross-section with different mechanisms
are summarized in a table~\ref{tab:P5XSec}. As we can see, 
the cross-section is largest for diagram $(a)$, when a pentaquark is produced
in a $P$-wave. Due to a node in the $r_{cc}$-dependence,  cross-section
($b$) is relatively small, and the numerical value depends a lot
on the position of the node, due to partial cancellation of small and
large-$r_{cc}$ contributions. The value chosen in the previous Section~\ref{sec:ModelWF}
corresponds to a node position in $\psi(2S)$ wave function known
from potential models. However, due to additional Van-der-Waals forces
acting on $\bar{c}c$ this position might be different. For diagram
($c$) the cross-section is small because the binding energies of
the initial proton inside a nucleus and light quarks inside a pentaquark
are small. The diagram ($d$) is small since a pentaquark is produced
diffractively and a proton has to interact both with light quarks
and $\bar{c}c$ pair separated by a large distance $\sim2$~fm.

\begin{figure}
\includegraphics[scale=0.8]{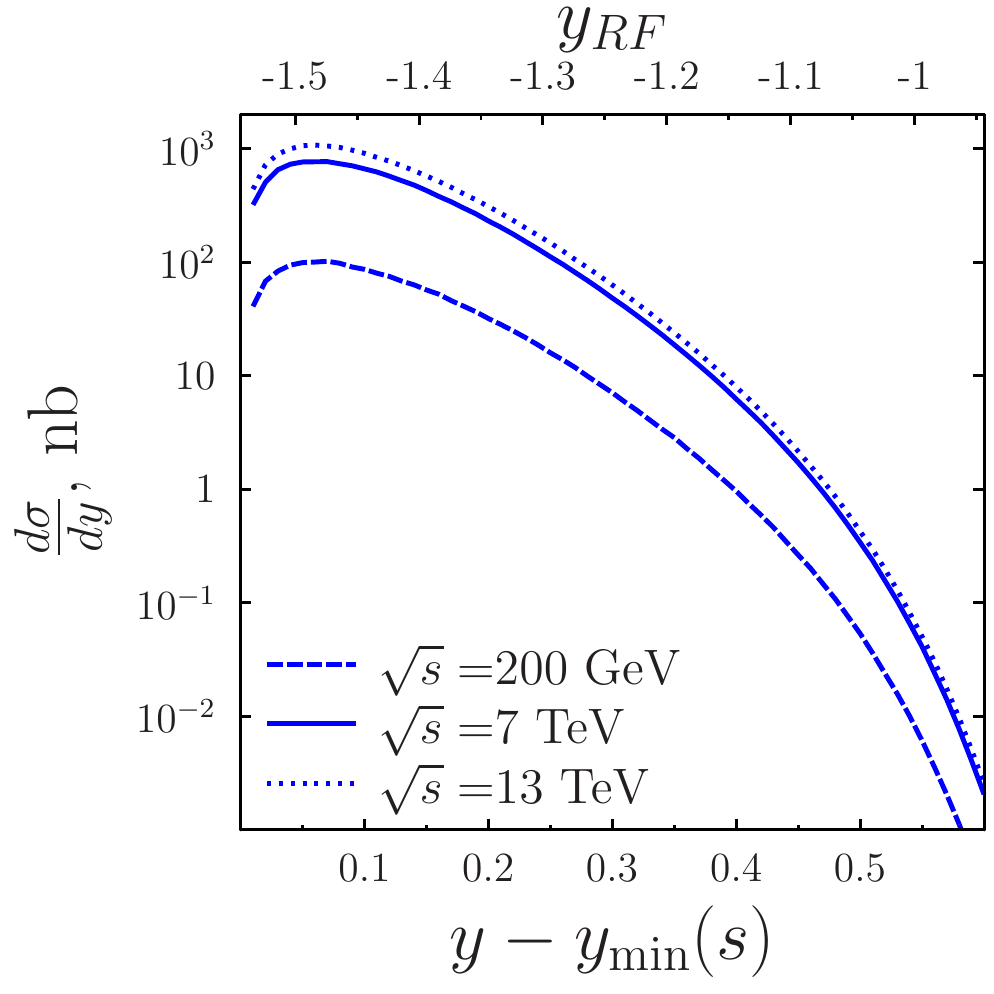}\includegraphics[scale=0.8]{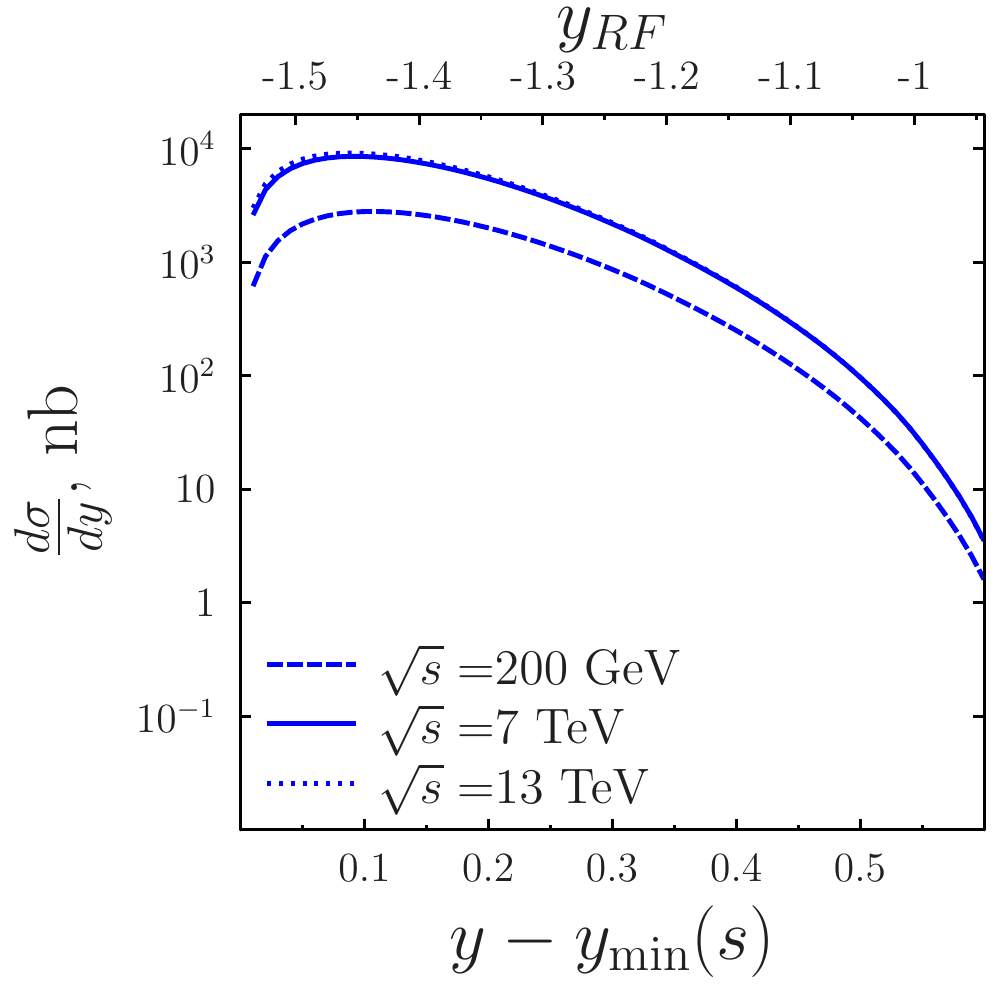}\protect\caption{\label{fig:sigma_y} Rapidity dependence of the cross-section $d\sigma/dy$
(if the pentaquark has a $\bar{c}c$ pair in the color singlet $S$-wave
(left) and $P$-wave (right)). $y$ is a pentaquark rapidity in a
nucleon-nucleon center of mass~(\ref{eq:rapidity}), $y_{RF}$ is
a rapidity in the nucleus rest frame~(\ref{eq:rapidity_RF}).}
\end{figure}

The rapidity dependence for the case of color singlet $\bar{c}c$
pair in $P_{c}$ is shown in the Figure~\ref{fig:sigma_y}. The decrease
of the curve at a small $y-y_{min}(s)$ happens due to a small-$x$
suppression of the overlap $f_{3,5}\left(x_{1},\alpha_{c}\approx\frac{1}{2}\right)\sim x_{1}$,
which implies that the $\bar{c}c$ pair and the proton separated by a
large rapidity gap cannot form a bound state. At large $y$ we have
a decrease due to suppression in the gluon PDF, which in the limit $x_{1}\to1$
behaves as $\sim\left(1-x_{1}\right)^{5}$. Due to a factorized form
of the pentaquark wave function~(\ref{eq:Pc_WF-sym}), the rapidity
dependence is similar for all cross-sections.

\begin{table}
\begin{tabular}{|c|c|c|c|c|c|c|c|}
\hline 
$\sqrt{s}$ & $y_{min}\left(\sqrt{s},\, P_{c}^{\perp}\approx0\right)$ & $(a)$ & $(b)$ & $(c),\, L=0$ & $(c),\, L=1$ & $(d),\, L=0$ & $(d),\, L=1$\tabularnewline
\hline 
$200$~GeV & 3.8 & $0.6\,{\rm \mu b}$ & $16\,{\rm nb}$ & $6.5\,{\rm nb}$ & $47\,{\rm nb}$ & $2.9\,{\rm nb}$ & $8.9\,{\rm nb}$\tabularnewline
\hline 
$7$~TeV & 7.4 & $1.9\,{\rm \mu b}$ & $120\,{\rm nb}$ & $137\,{\rm nb}$ & $135\,{\rm nb}$ & $19\,{\rm nb}$ & $8.2\,{\rm nb}$\tabularnewline
\hline 
$13$~TeV & 8 & $2\,{\rm \mu b}$ & $163\,{\rm nb}$ & $208\,{\rm nb}$ & $142\,{\rm nb}$ & $20.4\,{\rm nb}$ & $6.08\,{\rm nb}$\tabularnewline
\hline 
 &  &  &  &  &  &  & \tabularnewline
\hline 
\end{tabular}\protect\caption{\label{tab:P5XSec}Pentaquark production cross-sections with different
mechanisms. Diagrams $(a)$ and $(b)$ correspond to color singlet,
$(c)$ and $(d)$ to color octet states. $L$ is the orbital angular
momentum of the pentaquark. }
\end{table}

To understand the magnitude of the cross-section, we may for the moment
assume dominance of the lowest charmonium state with proper quantum
numbers ($\chi_{c}$ for a diagram ($a$) and $\psi(2S)$ for a diagram
($b$)), yielding 
\begin{equation}
\left(\frac{d\sigma^{(a)}}{dy_{P_{c}}}\right)_{{\rm estimate}}\sim\frac{1+x_{1}}{x_{1}}\,\mathcal{N}\, f_{3,5}^{2}\left(x_{1},\alpha_{c}\approx\frac{1}{2}\right)\frac{d\sigma^{(pA\to\chi_{c})}}{dy_{\chi_{c}}},\label{eq:Chic_est}
\end{equation}

\begin{equation}
\left(\frac{d\sigma^{(b)}}{dy_{P_{c}}}\right)_{{\rm estimate}}\sim\frac{1+x_{1}}{x_{1}}\,\mathcal{N}\, f_{3,5}^{2}\left(x_{1},\alpha_{c}\approx\frac{1}{2}\right)\frac{d\sigma^{(pA\to\psi(2S))}}{dy_{\psi(2S)}}.\label{eq:JPsi_est}
\end{equation}
where 

\begin{eqnarray}
f_{3,5}\left(x_{1},\alpha_{c}\right) & \equiv & \int\prod_{i=1}^{3}d\alpha_{i}f_{5}\left(\left\{ \frac{\alpha_{i}}{1+x_{1}}\right\} ,\,\frac{x_{1}\alpha_{c}}{1+x_{1}},\,\frac{x_{1}\bar{\alpha}_{c}}{1+x_{1}}\right)f_{3}\left(\alpha_{i}\right)\delta\left(\sum_{i}\alpha_{i}=1\right),
\end{eqnarray}
 is an overlap of the light cone distributions given by~\ref{eq:Brodsky_IC},
and 
\begin{equation}
\mathcal{N}=\left[\int d^{2}R_{cc}\Phi_{LC}^{2D}\left(-\frac{M_{P_{c}}}{M_{P_{c}}-2\, m_{c}}\vec{R}_{cc}\right)\sqrt{\frac{1}{\pi\left\langle R_{cc}^{2}\right\rangle }}\,\exp\left(-\frac{R_{\bar{c}c}^{2}}{\left\langle R_{cc}^{2}\right\rangle }\right)\right]^{2}
\end{equation}
is an overlap of the transverse parts of the wave functions. We took
$\alpha_{c}\approx1/2$ in~(\ref{eq:Chic_est},\ref{eq:JPsi_est})
since the momentum of the gluon is split approximately in equal proportion
in a $\bar{c}c$ pair. Using the rapidity dependence from~\cite{Aamodt:2011gj,LHCb:2012af,Chatrchyan:2011kc},
we may get curves similar to what is shown in Figure~\ref{fig:sigma_y}. 

\begin{figure}
\includegraphics[scale=0.5]{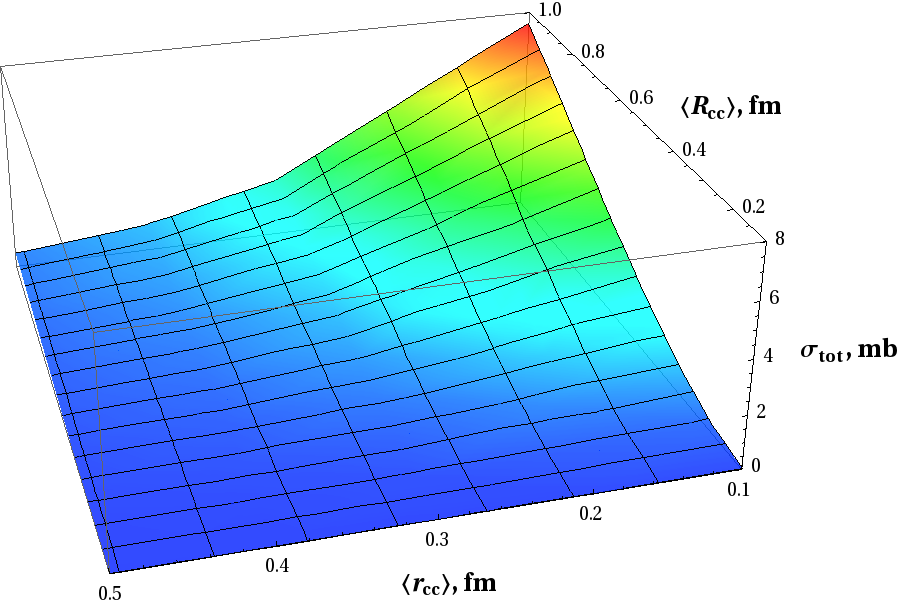}

\protect\caption{\label{fig:sigma-a}(color online) Dependence of the integrated cross-section~$\sigma^{(a)}$
(singlet $P$-wave $\bar{c}c$ pair) on pentaquark parameters $\left\langle r_{cc}\right\rangle $
and $\left\langle R_{cc}\right\rangle $ for $\sqrt{s}\approx7\,{\rm TeV}$.}
\end{figure}

In Figure~\ref{fig:sigma-a} we show the dependence of the
cross-section $\sigma_{a}$ on the pentaquark parameters $\left\langle r_{cc}\right\rangle $
and $\left\langle R_{cc}\right\rangle $, for $\sqrt{s}\approx7$~TeV.
A $\bar{c}c$ dipole produced from a gluon has a small size $\sim m_{c}^{-1}$,
so its overlap with a wide $r_{cc}$-distribution decreases when the
width of the distribution $\langle r_{cc}\rangle$ increases. Similar
origin has the dependence on $\left\langle R_{cc}\right\rangle $: the
overlap of the two-nucleon correlation function~(\ref{eq:WF_2D})
with the pentaquark wave function is small if the distance between $\bar{c}c$
and light quarks is zero, and is growing up to its peak value when
$\left\langle R_{cc}\right\rangle $ reaches a value of a few fm.
The true value of the parameter $\left\langle R_{cc}\right\rangle $
could be fixed in the future either from low-energy models of the
pentaquark structure or from phenomenological study of the $P_{\perp}$-slope
of the produced pentaquarks. Due to a factorized form of a parametrization~\ref{eq:Pc_WF-sym},
the dependence of the other cross-sections ($(b)-(d)$) on the parameters
$\left\langle R_{cc}\right\rangle $ and $\left\langle r_{cc}\right\rangle $
has a similar shape. 

\section{Conclusion}

In this paper we suggested that a hidden charm pentaquark $P_{c}^{+}$
might be produced in proton-nucleus collisions. In a collider kinematics,
this production happens at forward rapidities, whereas in the fixed
target experiments~\cite{Brodsky:2012vg,Lansberg:2012kf} (nucleus
rest frame) the pentaquarks are produced with relatively small rapidities
and can be easily detected via invariant mass analysis of possible
pentaquark decay products, as discussed in Sections~\ref{sec:Mechanism},~\ref{sec:NumericalResults}.
We estimated the cross-section for different color and internal orbital
momentum of the $\bar{c}c$ inside a pentaquark and find that it is sizable.
A key advantage of the suggested mechanism is that it does not involve
any electroweak intermediaries, which provides a larger cross-section
compared to photon-induced processes, and for typical $\mathcal{L}\sim{\rm fb}^{-1}$
integrated luminosities the total number of produced pentaquarks is
considerably larger than in case of weak decays of $\Lambda_{b}$
reported in an LHCb paper~\cite{Aaij:2015tga}. An additional appeal
of the suggested process is that it allows to access parameters of
pentaquark wave function. In particular, a rapidity distribution of
produced pentaquarks probes a fraction of pentaquark light-cone momentum
carried by $\bar{c}c$ pair. A slope of the $P_{T}$-distribution
is controlled by an average distance between center of mass of $P_{c}^{+}$
and center of the $\bar{c}c$ pair. 

We think that experimentalists should analyze mass distributions of possible
decay products of $P_{c}^{+}$($J/\psi+p$, $\chi_{c}p$, $\bar{D}$
and charmed baryon) in order to study its properties. If $P_{c}^{+}$
has neutral ``siblings'' with structure $udd\bar{c}c$ as suggested
by several models~\cite{Li:2015gta,Briceno:2015rlt,Feijoo:2015kts},
these could be also produced via $\bar{c}c+n\to P_{c}^{0}$ subprocess
in $p\, A$ collisions. In view of isospin invariance of strong interactions,
the cross-section of such process is related to a $P_{c}^{+}$ production
cross-section by a factor $(A-Z)/Z$.

When this manuscript was over, a similar process was suggested in~\cite{Wang:2016vxa},
in which a pentaquark $P_{c}^{+}$ is produced via coalescence of
quarks in $pA$ and $AA$ collisions. The evaluation was performed
at midrapidity region in center-of-mass system, when a contribution
from the so-called net quarks is negligible. 
% As was found in~\cite{Wang:2016vxa},
% the mechanism leads to relatively small yields for $P_{c}^{+}$. 
In contrast, in our mechanism %gives larger cross-sections since 
all the light quarks needed for the process stem from the nucleon. Since the
two mechanisms work in different rapidity intervals, a direct comparison
between them is not possible.

\section*{Acknowledgements}

We thank our colleagues at UTFSM university for encouraging discussions.
Especially we would like to thank E. Levin and B. Kopeliovich for
numerous pedagogical discussions and seminal critics. This research
was partially supported by the Fondecyt (Chile) grants 1140390 and
1140377. Powered@NLHPC: This research was partially supported by the
supercomputing infrastructure of the NLHPC (ECM-02). Also, we thank
Yuri Ivanov for technical support of the USM HPC cluster where part
of evaluations were done.


\begin{thebibliography}{10}
 \bibitem{Aaij:2015tga}R.~Aaij \emph{et al.} [LHCb Collaboration],
 Phys.~Rev.~Lett. \textbf{115}, 072001 (2015) [arXiv:1507.03414
 [hep-ex]].
 
 \bibitem{Golec-Biernat:2015aza}K.~Golec-Biernat, E.~Lewandowska,
 M.~Serino, Z.~Snyder and A.~M.~Stasto, arXiv:1507.08583 [hep-ph].
 
 \bibitem{GellMann:1964nj}M.~Gell-Mann, Phys.~Lett.~\textbf{8}
 (1964) 214.
 
 \bibitem{Brodsky:1997yr}S.~J.~Brodsky and F.~S.~Navarra, Phys.~Lett.~B
 \textbf{411}, 152 (1997) [hep-ph/9704348].
 
 \bibitem{Brodsky:1989jd}S.~J.~Brodsky, I.~A.~Schmidt and G.~F.~de
 Teramond, Phys.~Rev.~Lett. \textbf{64} (1990) 1011.
 
 \bibitem{Brodsky:1980pb}S.~J.~Brodsky, P.~Hoyer, C.~Peterson
 and N.~Sakai, Phys.~Lett.~B \textbf{93}, 451 (1980).
 
 \bibitem{Brodsky:1981se}S.~J.~Brodsky, C.~Peterson and N.~Sakai,
 Phys.~Rev.~D \textbf{23}, 2745 (1981).
 
 \bibitem{Karliner:2015ina}M.~Karliner and J.~L.~Rosner, Phys.~Rev.~Lett.~\textbf{115},
 (2015), 122001 [arXiv:1506.06386 [hep-ph]].
 
 \bibitem{Chen:2015moa}H.~X.~Chen, W.~Chen, X.~Liu, T.~G.~Steele
 and S.~L.~Zhu, Phys.~Rev.~Lett.~\textbf{115}, no. 17, 172001
 (2015) [arXiv:1507.03717 [hep-ph]].
 
 \bibitem{Wang:2015qlf}G.~J.~Wang, L.~Ma, X.~Liu and S.~L.~Zhu,
 arXiv:1511.04845 [hep-ph].
 
 \bibitem{He:2015cea}J.~He, arXiv:1507.05200 [hep-ph].
 
 \bibitem{Chen:2015loa}R.~Chen, X.~Liu, X.~Q.~Li and S.~L.~Zhu,
 Phys.~Rev.~Lett.~\textbf{115}, no. 13, 132002 (2015) [arXiv:1507.03704
 [hep-ph]].
 
 \bibitem{Scoccola:2015nia}N.~N.~Scoccola, D.~O.~Riska and M.~Rho,
 Phys.~Rev.~D \textbf{92}, no. 5, 051501 (2015) [arXiv:1508.01172
 [hep-ph]].
 
 \bibitem{Yang:2015bmv}G.~Yang and J.~Ping, arXiv:1511.09053 [hep-ph].
 
 \bibitem{Chen:2016heh}R.~Chen, X.~Liu and S.~L.~Zhu, arXiv:1601.03233
 [hep-ph].
 
 
 \bibitem{Meissner:2015mza}U.~G.~Meißner and J.~A.~Oller, Phys.~Lett.~B
 \textbf{751}, 59 (2015) [arXiv:1507.07478 [hep-ph]].
 
 \bibitem{Kahana:2015tkb}D.~E.~Kahana and S.~H.~Kahana, arXiv:1512.01902
 [hep-ph].
 
 \bibitem{Eides:2015dtr}M.~I.~Eides, V.~Y.~Petrov and M.~V.~Polyakov,
 arXiv:1512.00426 [hep-ph].
 
 \bibitem{Mironov:2015ica}A.~Mironov and A.~Morozov, JETP Lett.~\textbf{102},
 no. 5, 271 (2015) [arXiv:1507.04694 [hep-ph]].
 
 \bibitem{Xiao:2015fia}C.~W.~Xiao and U.-G.~Meißner, arXiv:1508.00924
 [hep-ph].
 
 \bibitem{Guo:2015umn}F.~K.~Guo, U.~G.~Meißner, W.~Wang and Z.~Yang,
 Phys.~Rev.~D \textbf{92}, no. 7, 071502 (2015) [arXiv:1507.04950
 [hep-ph]].
 
 \bibitem{Anisovich:2015xja}V.~V.~Anisovich, M.~A.~Matveev, A.~V.~Sarantsev
 and A.~N.~Semenova, Mod.~Phys.~Lett.~A \textbf{30}, 1550212 (2015)
 [arXiv:1509.03028 [hep-ph]].
 
 \bibitem{Wang:2015pcn}E.~Wang, H.~X.~Chen, L.~S.~Geng, D.~M.~Li
 and E.~Oset, arXiv:1512.01959 [hep-ph].
 
 \bibitem{Jimenez-Delgado:2014zga}P.~Jimenez-Delgado, T.~J.~Hobbs,
 J.~T.~Londergan and W.~Melnitchouk, Phys.~Rev.~Lett.~\textbf{114},
 no. 8, 082002 (2015) [arXiv:1408.1708 [hep-ph]].
 
 \bibitem{Li:2015gta}G.~N.~Li, M.~He and X.~G.~He, arXiv:1507.08252
 [hep-ph].
 
 \bibitem{Briceno:2015rlt}R.~A.~Briceno \emph{et al.}, arXiv:1511.06779
 [hep-ph].
 
 \bibitem{Feijoo:2015kts}A.~Feijoo, V.~K.~Magas, A.~Ramos and
 E.~Oset, arXiv:1512.08152 [hep-ph].
 
 \bibitem{Karliner:2015voa}M.~Karliner and J.~L.~Rosner, Phys.~Lett.~B
 \textbf{752}, 329 (2016) [arXiv:1508.01496 [hep-ph]].
 
 \bibitem{Kubarovsky:2015aaa}V.~Kubarovsky and M.~B.~Voloshin,
 Phys.~Rev.~D \textbf{92}, no. 3, 031502 (2015) [arXiv:1508.00888
 [hep-ph]].
 
 \bibitem{Wang:2015jsa}Q.~Wang, X.~H.~Liu and Q.~Zhao, Phys.~Rev.~D
 \textbf{92}, 034022 (2015) [arXiv:1508.00339 [hep-ph]].
 
 \bibitem{Lu:2015fva}Q.~F.~Lü, X.~Y.~Wang, J.~J.~Xie, X.~R.~Chen
 and Y.~B.~Dong, arXiv:1510.06271 [hep-ph].
 
 \bibitem{Hicks:2005gp}K.~H.~Hicks, Prog.~Part.~Nucl.~Phys.~\textbf{55},
 647 (2005) [hep-ex/0504027].
 
 \bibitem{Brambilla:2004wf}N.~Brambilla \emph{et al.} [Quarkonium
 Working Group Collaboration], hep-ph/0412158.
 
 \bibitem{Bedjidian:2004gd}M.~Bedjidian \emph{et al.}, hep-ph/0311048.
 
 \bibitem{Brodsky:2009cf}S.~J.~Brodsky and J.~P.~Lansberg, Phys.~Rev.~D
 \textbf{81}, 051502 (2010) [arXiv:0908.0754 [hep-ph]].
 
 \bibitem{Kopeliovich:2001ee}B.~Kopeliovich, A.~Tarasov and J.~Hufner,
 Nucl.~Phys.~A \textbf{696}, 669 (2001) [hep-ph/0104256].
 
 \bibitem{Kopeliovich:2005us}B.~Z.~Kopeliovich, I.~K.~Potashnikova
 and I.~Schmidt, Phys.~Rev.~C \textbf{73}, 034901 (2006) [hep-ph/0508277].
 
 \bibitem{Kopeliovich:2015qna}B.~Z.~Kopeliovich, E.~Levin, I.~Schmidt
 and M.~Siddikov, Phys.~Rev.~D \textbf{92}, no. 3, 034023 (2015)
 [arXiv:1501.01607 [hep-ph]].
 
 \bibitem{Kopeliovich:2015vqa}V.~Kopeliovich and I.~Potashnikova,
 arXiv:1510.05958 [hep-ph].
 
 \bibitem{Kopeliovich:2002yv}B.~Z.~Kopeliovich and A.~V.~Tarasov,
 Nucl.~Phys.~A \textbf{710}, 180 (2002) [hep-ph/0205151].
 
 \bibitem{Baranov:2007dw}S.~P.~Baranov and A.~Szczurek, Phys.~Rev.~D
 \textbf{77}, 054016 (2008) [arXiv:0710.1792 [hep-ph]].
 
 \bibitem{Baranov:2015laa}S.~P.~Baranov, A.~V.~Lipatov and N.~P.~Zotov,
 Eur.~Phys.~J.~C \textbf{75}, no. 9, 455 (2015) [arXiv:1508.05480
 [hep-ph]].
 
 \bibitem{Baranov:2015yea}S.~P.~Baranov, A.~V.~Lipatov and N.~P.~Zotov,
 arXiv:1510.02411 [hep-ph].
 
 \bibitem{GolecBiernat:1999ib}K.~J.~Golec-Biernat, A.~D.~Martin
 and M.~G.~Ryskin, Phys.~Lett.~B \textbf{456}, 232 (1999) [hep-ph/9903327].
 
 \bibitem{Kogut:1969xa}J.~B.~Kogut and D.~E.~Soper, Phys.~Rev.~D
 \textbf{1}, 2901 (1970).
 
 \bibitem{Korchemsky:2001nx}G.~P.~Korchemsky, J.~Kotanski and A.~N.~Manashov,
 Phys.~Rev.~Lett.~\textbf{88}, 122002 (2002) [hep-ph/0111185].
 
 \bibitem{ELBook}Yu. Kovchegov, E. Levin, ``\emph{Quantum Chromodynamics
 at high energy}'', Cambridge University Press, Cambridge, UK. ISBN:
 9780521112574
 
 \bibitem{Kowalski:2003hm}H.~Kowalski and D.~Teaney, Phys.~Rev.~D
 \textbf{68}, 114005 (2003) [hep-ph/0304189].
 
 \bibitem{Tagami:1959}T.~Tagami, Prog. Theor. Phys. \textbf{21} (4)
 533-561 (1959). [doi:10.1143/PTP.21.533]
 
 \bibitem{Levinger:1979jf}J.~S.~Levinger, Phys.~Lett. B \textbf{82},
 181 (1979).
 
 \bibitem{Alvioli:2009ab}M.~Alvioli, H.-J.~Drescher and M.~Strikman,
 Phys.~Lett.~B \textbf{680}, 225 (2009) [arXiv:0905.2670 [nucl-th]].
 
 \bibitem{Colle:2013nna}C.~Colle, W.~Cosyn, J.~Ryckebusch and M.~Vanhalst,
 Phys.~Rev.~C \textbf{89}, no. 2, 024603 (2014) [arXiv:1311.1980
 [nucl-th]].
 
 \bibitem{Viollier:1976ab}R.~D.~Viollier and J.~D.~Walecka, Acta
 Phys.~Polon.~B \textbf{8}, 25 (1977).
 
 \bibitem{Cao:2012cv}X.~G.~Cao, X.~Z.~Cai, Y.~G.~Ma, D.~Q.~Fang,
 G.~Q.~Zhang, W.~Guo, J.~G.~Chen and J.~S.~Wang, Phys.~Rev.~C
 \textbf{86}, 044620 (2012) [arXiv:1211.6906 [nucl-th]].
 
 \bibitem{Neff:2015xda}T.~Neff, H.~Feldmeier and W.~Horiuchi, Phys.~Rev.~C
 \textbf{92}, no. 2, 024003 (2015) [arXiv:1506.02237 [nucl-th]].
 
 \bibitem{Egiyan:2005hs}K.~S.~Egiyan \emph{et al.} [CLAS Collaboration],
 Phys.~Rev.~Lett.~\textbf{96}, 082501 (2006) [nucl-ex/0508026].
 
 \bibitem{Frankfurt:1993sp}L.~L.~Frankfurt, M.~I.~Strikman, D.~B.~Day
 and M.~Sargsian, Phys.~Rev.~C \textbf{48}, 2451 (1993).
 
 \bibitem{Piasetzky:2006ai}E.~Piasetzky, M.~Sargsian, L.~Frankfurt,
 M.~Strikman and J.~W.~Watson, Phys.~Rev.~Lett.~\textbf{97},
 162504 (2006) [nucl-th/0604012].
 
 \bibitem{Hen:2014nza}O.~Hen \emph{et al.}, Science \textbf{346},
 614 (2014)  [arXiv:1412.0138 [nucl-ex]].
 
 \bibitem{Colle:2015ena}C.~Colle, O.~Hen, W.~Cosyn, I.~Korover,
 E.~Piasetzky, J.~Ryckebusch and L.~B.~Weinstein, Phys.~Rev.~C
 \textbf{92}, no. 2, 024604 (2015) [arXiv:1503.06050 [nucl-th]].
 
 \bibitem{Alvioli:2008rw}M.~Alvioli, C.~Ciofi degli Atti, I.~Marchino,
 V.~Palli and H.~Morita, Phys.~Rev.~C \textbf{78}, 031601 (2008)
 [arXiv:0807.0873 [nucl-th]].
 
 \bibitem{ArgonneWF}R.~B.\textbf{~}Wiringa, V.~G.~J.~Stoks and
 R.~Schiavilla, Phys.~Rev.~C \textbf{51} (1995) 38. 
 
 \bibitem{Frankfurt:1997fj}L.~Frankfurt, W.~Koepf and M.~Strikman,
 Phys.~Rev.~D \textbf{57}, 512 (1998) [hep-ph/9702216].
 
 \bibitem{Hoyer:1999xe}P.~Hoyer and S.~Peigne, Phys.~Rev.~D \textbf{61},
 031501 (2000) [hep-ph/9909519].
 
 \bibitem{Diakonov:1997mm}D.~Diakonov, V.~Petrov and M.~V.~Polyakov,
 Z.~Phys.~A \textbf{359}, 305 (1997) [hep-ph/9703373].
 
 \bibitem{Lorce:2006nq}C.~Lorce, Phys.~Rev.~D \textbf{74}, 054019
 (2006) [hep-ph/0603231].
 
 \bibitem{Eichten:1979ms}E.~Eichten, K.~Gottfried, T.~Kinoshita,
 K.~D.~Lane and T.~M.~Yan, Phys.~Rev.~D \textbf{21} (1980) 203.
 
 \bibitem{Rezaeian:2012ji}A.~H.~Rezaeian, M.~Siddikov, M.~Van
 de Klundert and R.~Venugopalan, Phys.~Rev.~D \textbf{87}, no. 3,
 034002 (2013) [arXiv:1212.2974].
 
 \bibitem{Aamodt:2011gj}K.~Aamodt \emph{et al.} [ALICE Collaboration],
 Phys.~Lett.~B 704, 442 (2011) [Phys.~Lett. B \textbf{718}, 692
 (2012)] [arXiv:1105.0380 [hep-ex]].
 
 \bibitem{LHCb:2012af}R.~Aaij \emph{et al.} [LHCb Collaboration],
 Phys.~Lett.~B \textbf{718}, 431 (2013) [arXiv:1204.1462 [hep-ex]].
 
 \bibitem{Chatrchyan:2011kc}S.~Chatrchyan \emph{et al.} [CMS Collaboration],
 JHEP \textbf{1202}, 011 (2012) [arXiv:1111.1557 [hep-ex]].
 
 \bibitem{Brodsky:2012vg}S.~J.~Brodsky, F.~Fleuret, C.~Hadjidakis
 and J.~P.~Lansberg, Phys.~Rept.~\textbf{522} (2013), 239 [arXiv:1202.6585
 [hep-ph]].
 
 \bibitem{Lansberg:2012kf}J.~P.~Lansberg, S.~J.~Brodsky, F.~Fleuret
 and C.~Hadjidakis, Few Body Syst.~\textbf{53}, 11 (2012) [arXiv:1204.5793
 [hep-ph]].
 
 \bibitem{Wang:2016vxa}R.~Q.~Wang, J.~Song, K.~J.~Sun, L.~W.~Chen,
 G.~Li and F.~L.~Shao, arXiv:1601.02835 [hep-ph].
 
\end{thebibliography}
 \end{document}